\documentclass[12pt]{article}
\pdfoutput=1
\usepackage{jheppub}
\usepackage{simplewick}
\usepackage{verbatim}
\usepackage{graphics, color}
\usepackage{graphicx}
\usepackage{amsmath}
\usepackage{amssymb}
\usepackage{amsthm}
\usepackage{amsfonts}
\usepackage[usenames,dvipsnames]{xcolor}
\usepackage[normalem]{ulem}
\usepackage{dsfont}
\usepackage{mathabx}
\usepackage{enumitem}
\usepackage[english]{babel}
\usepackage{slashed}

\DeclareMathOperator{\Tr}{Tr}
\renewcommand{\Re}{\operatorname{Re}}

\newcommand{\RR}{\mathbb{R}}

\newcommand{\ZZ}{\mathbb{Z}}
\renewcommand{\sl}{\mathfrak{sl}(2)}

\newcommand{\lan}{\langle}
\newcommand{\ran}{\rangle}

\renewcommand{\ln}{\log}

\definecolor{grey}{rgb}{.5,.5,.5}
\definecolor{bluegreen}{rgb}{0,.5,.5}
\definecolor{darkgreen}{rgb}{0,.5,0}

\begin{document}

\title{The entropy of bulk quantum fields and the entanglement wedge of an evaporating black hole}
\author[a]{Ahmed Almheiri,}
\author[b,c]{Netta Engelhardt,}
\author[d]{Donald Marolf,}
\author[d]{Henry Maxfield}
\affiliation[a]{Institute for Advanced Study,  Princeton, NJ 08540, USA}
\affiliation[b]{Department of Physics, Princeton University, Princeton, NJ 08544, USA}
\affiliation[c]{Gravity Initiative, Princeton University, Princeton NJ 08544, USA}
\affiliation[d]{Physics Department, University of California, Santa Barbara, CA 93016}
\emailAdd{almheiri@ias.edu}
\emailAdd{nengelhardt@princeton.edu}
\emailAdd{marolf@ucsb.edu}
\emailAdd{hmaxfield@physics.ucsb.edu}
\abstract{Bulk quantum fields are often said to contribute to the generalized entropy $\frac{A}{4G_N} +S_{\mathrm{bulk}}$ only at $O(1)$. Nonetheless, in the context of evaporating black holes, $O(1/G_N)$  gradients in  $S_{\mathrm{bulk}}$ can arise due to large boosts, introducing a quantum extremal surface far from any classical extremal surface. We examine the effect of such bulk quantum effects on quantum extremal surfaces (QESs) and the resulting entanglement wedge in a simple two-boundary $2d$ bulk system defined by Jackiw-Teitelboim gravity coupled to a 1+1 CFT.  Turning on a coupling between one boundary and a further external auxiliary system which functions as a heat sink allows a two-sided otherwise-eternal black hole to evaporate on one side.
We find the generalized entropy of the QES to behave as expected from general considerations of unitarity, and in particular that ingoing information disappears from the entanglement wedge after a scambling time $\frac{\beta}{2\pi} \ln \Delta S + O(1)$ in accord with expectations for holographic implementations of the Hayden-Preskill protocol.  We also find an interesting QES phase transition at what one might call the Page time for our process.}

\maketitle

\section{Introduction}

A key concept in our current understanding of holographic dualities is the entanglement wedge $W_A$ of a bulk spacetime associated with a given region $A$ in the dual holographic field theory \cite{Czech:2012bh,Wall:2012uf,Headrick:2014cta}.  At the level of classical bulk physics, $W_A$ is obtained by first constructing the associated Hubeny-Rangamani-Takayanagi (HRT) surface \cite{Hubeny:2007xt}, which is the minimal-area codimension-2 bulk extremal surface homologous to $A$ in an appropriate sense \cite{Headrick:2007km}.  The physics in the wedge $W_A$ can then be reconstructed from field theory degrees of freedom in $A$ \cite{Almheiri:2014lwa,Jafferis:2015del,Dong:2016eik,Faulkner:2017vdd}.

Key steps in the above arguments rely on identifying the area ${\cal A}$ of the HRT surface as $4G_N$ times the von Neumann entropy of the field theory degrees of freedom in $A$ \cite{Ryu:2006bv,Ryu:2006ef,Hubeny:2007xt,Lewkowycz:2013nqa,Dong:2016hjy}.  Here $G_N$ is the bulk Newton constant and the bulk is treated classically.  But at the quantum level the von Neumann entropy of the field theory degrees of freedom in $A$ is instead the so-called generalized entropy of a bulk surface $X$ \cite{Faulkner:2013ana}, which for Einstein-Hilbert gravity coupled to $O(1)$ bulk quantum fields may be written $S_{\mathrm{gen}} = \frac{{\cal A}(X)}{4G_N} + S_{\mathrm{bulk}}(X)$, where $S_{\mathrm{bulk}}(X)$ is a von Neumann entropy of bulk quantum fields on one side of $X$.  As explained in \cite{Engelhardt:2014gca} and partially verified in \cite{Dong:2016hjy}, this means that bulk quantum effects should move the boundary of $W_A$ to a so-called quantum extremal surface $X$ extremizing
$S_{\mathrm{gen}}$ as defined by ${\cal A}[X]$ and by the von Neumann entropy of bulk quantum fields between $X$ and $A$.  As before, $X$ should satisfy the homology constraint and, when there is more than one such quantum extremal surface, we should choose the one minimizing $S_{\mathrm{gen}}$.

One often thinks of such quantum corrections as being small.  In many contexts $S_{\mathrm{bulk}}$ is indeed $O(1)$ and the relevant quantum and classical extremal surfaces nearly coincide.  In such contexts one should also consider corrections to the classical formula $\frac{{\cal A}}{4G_N}$ associated with higher derivative corrections to the gravitational effective action \cite{Wald:1993nt,Iyer:1995kg,Jacobson:1993vj,Dong:2013qoa,Miao:2014nxa}.  But in contexts involving long times and/or long distances, secular effects can cause $S_{\mathrm{bulk}}$ to grow and in some cases to become of order $1/G_N$, and large boosts can cause sharp gradients even when $S_{\mathrm{bulk}}$ remains $O(1)$. The last of these, which we investigate here, is particularly natural in the context of black hole evaporation, where the semi-classical Hawking effect leads to bulk entropy comparable to the Bekenstein-Hawking entropy of the original black hole \cite{Zurek:1982zz,Page:1983ug} and large boosts arise naturally from time evolution.

\begin{figure}
\begin{center}
\includegraphics[height=4cm]{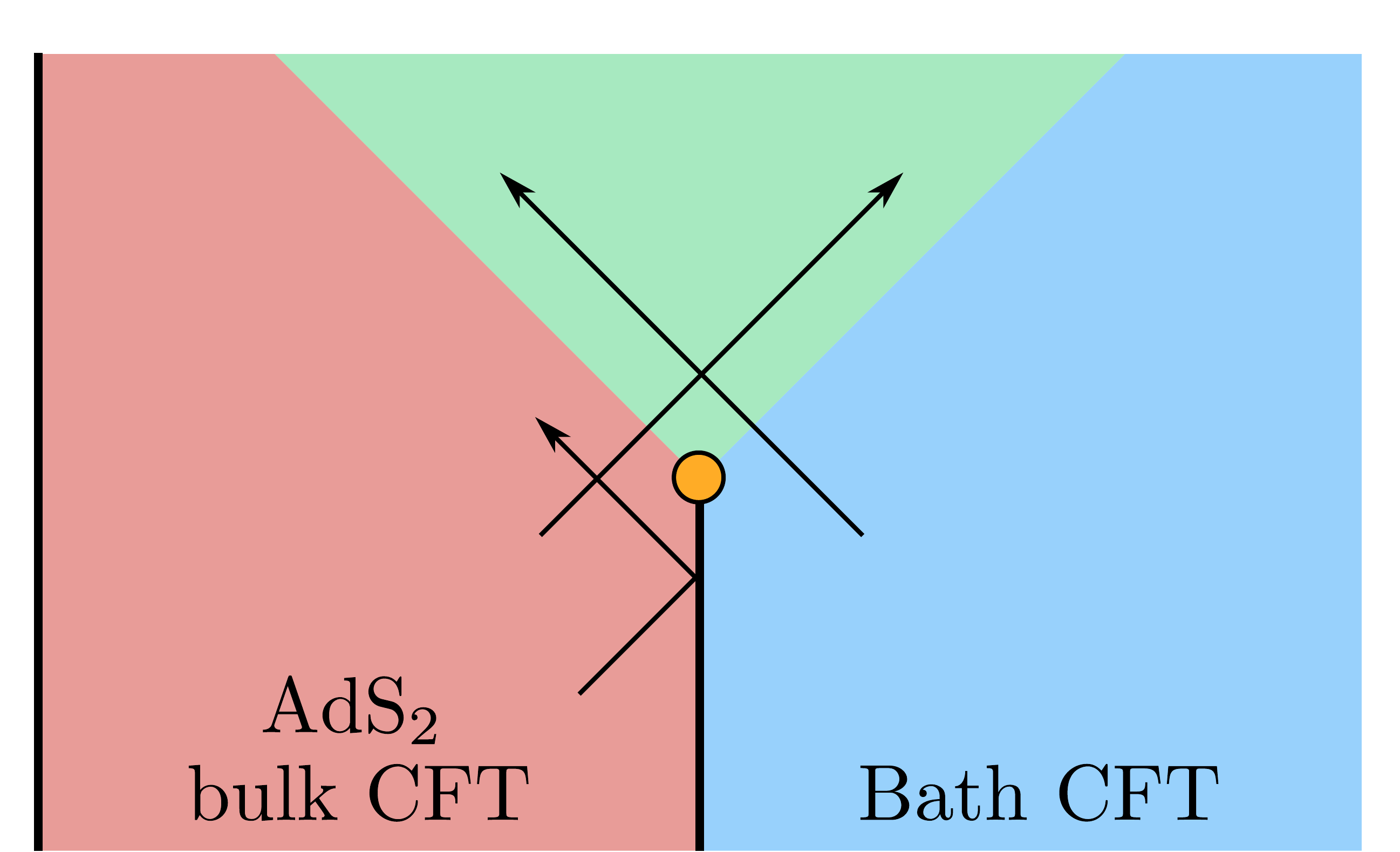}
\end{center}
\caption{Our two-sided AdS$_2$ system initially has reflecting boundary conditions (solid vertical lines) on its right boundary. An independent copy $B$ of our CFT on the right half of Minkowski space which will play the role of the bath also begins with reflecting boundary conditions. At some finite time (orange dot), the right-AdS$_2$ boundary conditions become transparent, coupling the AdS$_2$ CFT to the Bath CFT.}
\label{fig:coupling}
\end{figure}

We study a simple model where such $O(1/G_N)$ gradients of bulk entropy can be calculated in detail
and the ensuing effects on quantum extremal surfaces (QESs) and entanglement wedges can be studied. We consider a standard two-sided AdS$_2$ black hole in Jackiw-Teitelboim (JT) gravity coupled to a 1+1 CFT in the Hartle-Hawking state.  With reflecting boundary conditions at AdS infinity, the state is independent of time.  But at a finite time we couple the right boundary of our system to an auxiliary system $B$, which functions as a bath, or heat sink.  We take $B$ to (1) be a copy of the same 1+1 CFT on the right half of Minkowski space and (2) begin in its own vacuum.  The coupling is such that, after a short transition, the right boundary is fully transparent.  In effect, the coupling simply glues the origin of the auxiliary Minkowski space to the right AdS$_2$ boundary; see figure \ref{fig:coupling}.    This leads to evaporation on the right of the two-sided AdS$_2$ black hole.

Because the AdS$_2$ system is no longer isolated, bulk von Neumann entropies depend on a choice of Cauchy slice, or at least a choice of where such slices meet the right AdS$_2$ boundary.  In this sense, the QES of the right boundary becomes time-dependent. The time-dependent QESs may be viewed as a proxy for what one would find if one turned off the coupling at the given time, used the data on the stated Cauchy surface as initial data for a new AdS$_2$ bulk, and computed the QES in the resulting isolated spacetime.  The isolated QES and the proxy QES in the coupled spacetime will coincide up to corrections associated with the details of how the coupling is switched off.  This setting and aspects of JT gravity are reviewed in section \ref{sec:setting}, while section \ref{sec:matter} presents initial studies of the matter sector.

The proxy QESs are studied in section \ref{sec:QES}.
Although we consider only standard perturbative semiclassical bulk physics, tracking the proxy QES and computing $S_{\mathrm{gen}}$ as a function of boundary time reproduces features one would expect from general considerations of fully unitary evolution. In particular, the Page time, when the fine-grained von Neumann entropy of the black hole saturates at the coarse-grained thermodynamic entropy, is marked by a phase transition where the quantum extremal surface jumps. Thereafter, the location of the quantum extremal surface gives a quantum geometric realization of the Hayden-Preskill protocol \cite{Hayden:2007cs}, as described holographically in \cite{Almheiri:2018xdw}.  For the convenience of the reader, the technical results are then summarized in section \ref{sec:sum}.

An important part of the above description is the gap between the QESs $X_L$ and $X_R$ associated with the left and right boundaries of AdS$_2$, and this gap is discussed further in section \ref{sec:gap}.
We close with further discussion in section \ref{sec:disc}, which in particular describes analogous effects in cases where black holes evaporate more completely.  The final interpretation of such results is unclear, but will clearly fuel further discussion of black hole information puzzles, firewalls, state-dependence, and related issues; see e.g. \cite{Harlow:2014yka,Marolf:2017jkr} for recent reviews and \cite{Almheiri:2018xdw} for further recent discussion.

Note to reader: While this work was underway we learned that related results were independently found in the work \cite{Penington:2019npb} which will appear on the arxiv simultaneously.

\subsection{Holographic Hayden-Preskill}
\label{sec:HoloHP}

Before proceeding with the main paper in section \ref{sec:setting}, we pause to give a brief review of the Hayden-Preskill protocol \cite{Hayden:2007cs}, and expectations for its holographic realization \cite{Almheiri:2018xdw}. The protocol  considers an old black hole past the Page time which is maximally entangled with its early radiation. Assuming the black hole is governed by a sufficiently scrambling internal Hamiltonian, abstract quantum information reasoning is used to show that information thrown into this black hole would be recoverable from the radiation in a relatively short time compared to the black hole lifetime.

The protocol can be described as follows: Consider throwing some information $m$ in the state $| i \ran_m$ into an old black hole ${\cal B}$ maximally entangled with the early radiation ${\cal E}$ in the state $| \psi \ran_{{\cal BE}}$. After allowing for the black hole interior to scramble via its own internal unitary dynamics governed by some unitary $U_{m \cal{B}}$, the black hole is allowed to evaporate into some new radiation $\cal{L}$ with the remaining black hole given by $\cal{B}'$. This process is described by
\begin{align}
|i \ran_{m} | \psi \ran_{{\cal BE}} \rightarrow U_{m \cal{B}} | i \ran_m | \psi \ran_{{\cal BE}}  \equiv | \Psi_i \ran_{m{\cal BE}} = | \Psi_i \ran_{{\cal B' L E}}.
\end{align}
The last equality just comes from the identification of $m\cal{B}$ and ${\cal{B' L}}$. The result of \cite{Hayden:2007cs} is that it is sufficient that $U_{m \cal{B}}$ to be drawn from a unitary 2-design\footnote{It is defined as that which coincides with the Harr measure up to second moments in $U_{ij} U_{kl}^\dagger$.} for the message to be recoverable from the radiation subsystem $\cal{L E}$, in the sense that
\begin{align}
\forall i, \ \ \exists V_{{\cal LE}} \ \ \mathrm{s.t.} \  \ V_{{\cal LE}} | \Psi_i \ran_{{\cal B' L E}} = | i \ran_{l_1} \otimes | \chi \ran_{{\cal B}'l_2} ,\label{HPcondition}
\end{align}
for some fixed $| \chi \ran$, and where $l_1,2$ are some Hilbert space factors of $\cal{L E}$.

This protocol predicts an interesting time scale for when the message appears in the Hawking radiation, which stems from the assumption that the black hole internal unitary dynamics is given by a unitary 2-design. These circuits have logarithmic depth in the number of qubits of the black hole and thus naturally suggest the time scale
\begin{align}
t_{HP} \sim {1 \over T} \log S_{BH},
\end{align}
where $T$ and $S_{BH}$ are the temperature and entropy of the black hole respectively. This time, called the scrambling time, places a lower bound on the time needed before the message appears in the radiation. We will see in section \ref{sec:QES} how this timescale naturally arises  in a precise form from the evolution of the QES of an evaporating black hole.

\begin{figure}
\begin{center}
\includegraphics[height=5cm]{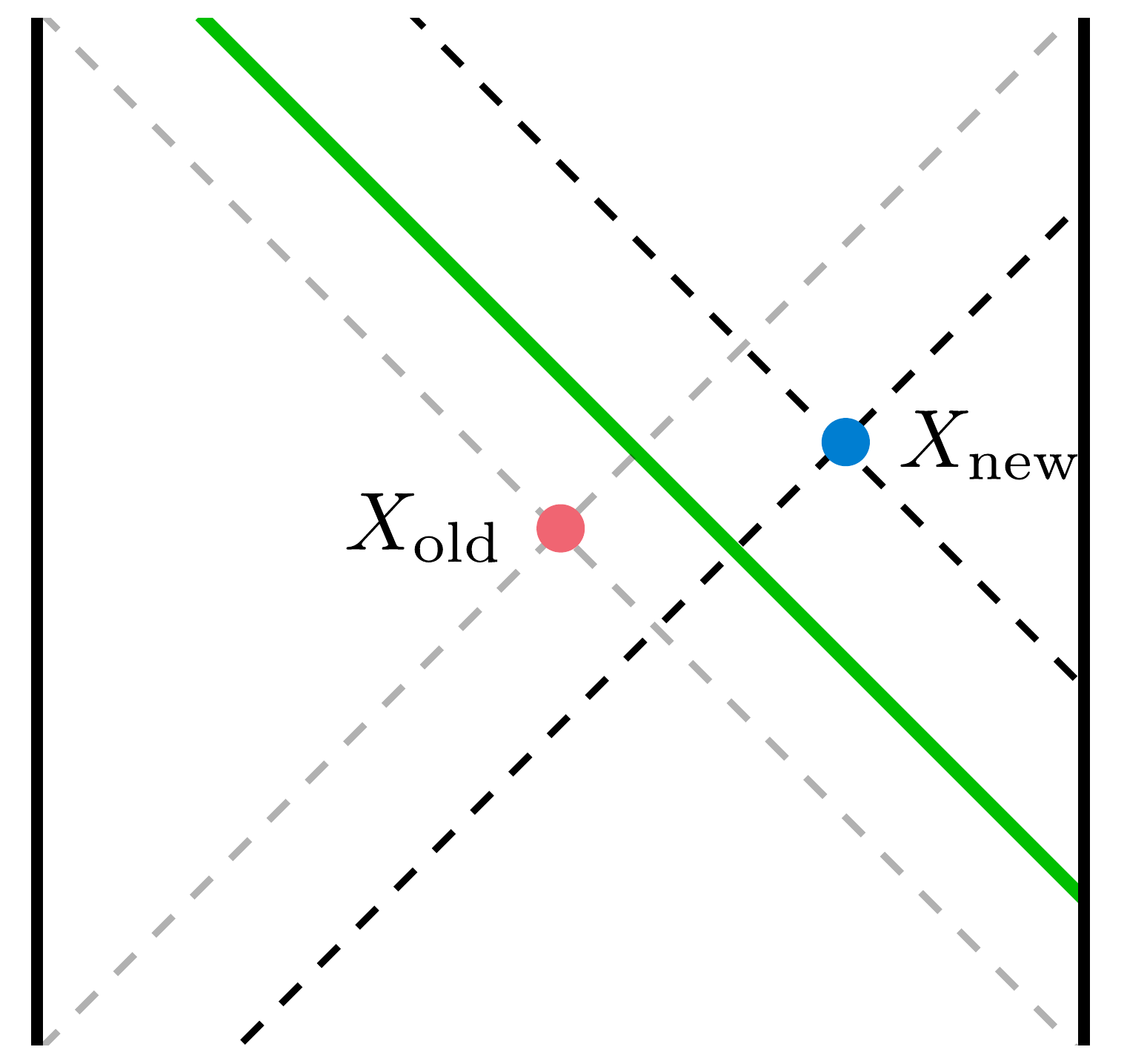}
\end{center}
\caption{After the evaporation of the right CFT $R$ into the left $L$, the quantum extremal surface moves in a spacelike direction towards the right boundary from $X_\mathrm{old}$ to $X_\mathrm{new}$. A message sent into $R$ in the past will escape the new entanglement wedge of $R$ and enter that of $L$.}
\label{fig:HPprotocol}
\end{figure}

As stated, the Hayden-Preskill protocol is a general statement about scrambling systems and would therefore naturally apply to the case of two entangled holographic CFTs. The setup of this protocol in the holographic context was recently discussed in \citep{Almheiri:2018xdw}, where the analogue of flat space post-Page time black hole is two entangled holographic CFTs,  $L \otimes R$,  in the thermofield double dual to the eternal black hole in AdS.

One obvious objection at this point is that the eternal black hole does not evaporate, so any information injected on one side will remain there eternally. We can mimic black hole evaporation by extracting energy from $R$ and dumping it into $L$; the right and left play the role of an old black hole and its early radiation respectively. Our goal is to study the evolution of the quantum extremal surface in the evaporating spacetime as a proxy for understanding, at each time, properties that the bulk-boundary dictionary would have if the coupling were turned off. Applying the Hayden-Preskill reasoning in this case would say that information sent into $R$ should become reconstructable in $L$ after a few scrambling times.

The general evaporation protocol will be as follows:
\begin{enumerate}
\item Start with two entangled CFTs in the thermofield double state with temperature above the Hawking-Page transition dual to the eternal black hole.
\item Introduce an auxiliary bath system $B$, taken to be a large system in the vacuum, which we couple to $R$ and allow the Hawking radiation to be extracted from $R$ to $B$. This is provided in AdS by imposing absorbing boundary conditions on the asymptotic boundary of $R$.
\item This Hawking radiation is then transferred into $L$. One can imagine transferring this information into any of the many low occupation modes on the left. This will excite the state of the quantum fields on the left exterior.
\end{enumerate}
The end result will be a new pure state of $L\otimes R$ where the ADM energy on the right is lower than that on the left (and lower than the initial energy on the right), with smaller entanglement between the two.

We will explain in detail in this paper how this protocol achieves Hayden-Preskill by inducing a motion of the QES surface away from the bifurcate horizon, causing the information injected into $R$ at early times to escape the entanglement wedge of that side. The rough idea is shown in figure \ref{fig:HPprotocol}. Since we expect that the modified state of $LR$ still exhibits complementary recovery, the inserted message enters the entanglement wedge of $L$, and can therefore be decoded from it, in the sense of \eqref{HPcondition}.

The scrambling time for near extremal black holes found in \cite{Leichenauer:2014nxa} is controlled by the ADM energy $E$ above the ground state and the energy of the perturbation $\delta E$ thrown into the black hole, assumed to be much smaller than $E$, as
\begin{align}
t_{scr} = \alpha_S {\beta \over 2 \pi} \ln {E \over \delta E}, \label{tscr1}
\end{align}
for some constant $\alpha_S$. This result indicates the Hayden-Preskill time should be
\begin{align}
t_{HP} = \alpha_{HP} {\beta \over 2 \pi} \ln (S - S_0), \label{thp1}
\end{align}
for a small message with $\delta E \sim E/(S-S_0)$, and for some constant $\alpha_{HP}$ which a priori could be distinct from $\alpha_S$. The realization of the Hayden-Preskill protocol in our context will be exhibited by the lag of the QES of the state at time $t$ by an amount $t_{HP}$ in null ingoing time; this would ensure that messages thrown in prior to $t - t_{HP}$ would escape the entanglement wedge. We confirm this expectation and determine the values of $\alpha_{HP}$ and $\alpha_{S}$ for systems dual to JT gravity.

We wish to emphasize the difference between this described protocol and the recent story of making a wormhole traversable via a double trace deformation \cite{Gao:2016bin, Maldacena:2017axo}. Traversability is achieved by violating the Averaged Null Energy Condition (ANEC) on the horizon, which provides a message falling into the horizon sufficent time advance that it emerges into the other asymptotic region of the eternal black hole. This is in contrast with the proposal of this paper where it is the evolution of the dictionary under the evaporation protocol that renders the message recoverable from the other boundary CFT.

Second, the traversable wormhole protocol takes advantage of the careful local correlations in the TFD between the two CFTs at $t = 0$ (or by boost invariance -  opposite times) and picks a deformation with large connected expectation value. This sensitivity to the state  implies a sensitivity to the time at which message is thrown in.  In particular, it works best for messages thrown in at around the scrambling time prior to turning on the interaction. A message sent in too early would spoil the delicate correlations in the TFD, thereby ruining the efficacy of the deformation. In the bulk, this is interpreted as the failure of the eikonal approximation of scattering between the message and the negative stress tensor, which precludes the necessary time advance for traverability \cite{Maldacena:2017axo}. Sent in late, the message simply doesn't get enough of a time advance to make it through. The evaporation protocol in this paper does not suffer from this issue, and as we will see, all messages thrown into the black hole will eventually appear in the entanglement wedge of the complement after a scrambling time.

\section{Evaporating Near-Extremal Black Holes in JT Gravity}
\label{sec:setting}

\subsection{Review of JT Gravity}

We will study the evolution of the minimal quantum extremal surface (QES) in an evaporating black hole in JT gravity coupled to conformal matter. The dynamics of this theory are governed by the Lorentzian action $I = I_0 + I_G + I_M$ with
\begin{align}
&I_0 = {\phi_0 \over 16 \pi G_N} \left[ \int_{\cal M} d^2 x \sqrt{-g} \  R  + 2\int_{\partial M}  K \right], \\
&I_G = {1\over 16 \pi G_N} \left[ \int_{\cal M} d^2 x \sqrt{-g} \  \phi (R + 2)  + 2\int_{\partial M} \  \phi_b  K \right],  \\
&I_M =  I_{CFT}[g].
\end{align}
The dynamics of this model are especially simple. The gravitational action can be thought of as the dimensional reduction of a higher dimensional theory describing the s-wave sector of the near horizon limit of near extremal black holes \cite{Maldacena:2016upp, Engelsoy:2016xyb, Almheiri:2016fws, Sachdev:2019bjn}. From this perspective, the area of the transverse space in the higher dimensional theory becomes the dilaton $ \phi_0 + \phi$, thereby implicitly imposing the restriction $\phi_0 \gg \phi$. The action $I_0$ is a purely topological term and provides the extremal entropy of the black hole ${\phi_0 / 4 G_N}$.

The remaining gravitational dynamics are governed by the action $I_G$, which is not topological because $\phi$ is dynamical. This action is easily solved by integrating out the dilaton along an imaginary contour which imposes the constraint on the spacetime to have constant negative curvature via the delta function
\begin{align}
\delta (R+2),
\end{align}
which requires that the two dimensional metric is locally AdS$_2$. In Poincar\'e coordinates this is
\begin{align}
ds^2 = {-dt^2 + dz^2 \over z^2} = - {4 dx^+ dx^- \over \left( x^+ - x^- \right)^2}, \ \ \ x^\pm = t \pm z.
\end{align}
As we review below, this AdS$_2$ space should be thought of as an `ambient' rigid space of which the actual physical spacetime is a patch \cite{Almheiri:2014cka, Maldacena:2016upp, Engelsoy:2016xyb, Jensen:2016pah}. Varying the action with respect to the metric $g$ yields the constraints and equation of motion that couple the bulk CFT to the dilaton:
\begin{subequations}
\begin{align}
2 \partial_{x^+} \partial_{x^-} \phi + {4 \over (x^+ - x^-)^2} \phi &= 16 \pi G_N T_{x^+ x^-}, \\
-{1 \over (x^+ - x^-)^2} \partial_{x^+} \left( (x^+ - x^-)^2 \partial_{x^+} \phi \right) &= 8 \pi G_N T_{x^+ x^+}, \\
-{1 \over (x^+ - x^-)^2} \partial_{x^-} \left( (x^+ - x^-)^2 \partial_{x^-} \phi \right) &= 8 \pi G_N T_{x^- x^-}. \label{eom}
\end{align}
\end{subequations}
We work in the limit where the gravitational sector can be treated semiclassically, so we may replace the stress tensors with their expectation values, $T_{a b} = \lan T_{ab} \ran$.

It is often convenient to express JT gravity as the dynamics of the so-called `boundary particle' \cite{ Maldacena:2016upp, Engelsoy:2016xyb, Jensen:2016pah}. This is simply a reparametrization between the bulk Poincar\'e time near the boundary $t$ and the physical boundary time $u$. The location of this physical boundary is specified by the boundary condition on the bulk fields
\begin{align}\label{eq:guu}
g_{u u}\big|_{bdy}  = {1 \over \epsilon^2} = {- t'^2 + z'^2 \over z^2}, \
\phi = \phi_b = {\bar{\phi}_r \over \epsilon},
\end{align}
where $g_{uu}$ is the time-time component of the metric near the boundary along the physical boundary time $u$. The last equality in~\eqref{eq:guu} indicates that we are interested in large $\phi_{b}$ ($\phi_{b}\sim 1/\epsilon$) with fixed constant coefficient $\bar{\phi}_{r}$. With this choice, the JT action reduces to a boundary term given by
\begin{align}
S_G =  {1 \over 8 \pi G_N}\int_{\partial M} \  \phi_b  K  \rightarrow {\bar{\phi}_r \over 8 \pi G_N} \int du \{ f(u), u \},
\end{align}
where $t=f(u)$ is a diffeomorphism giving Poincar\'e time $t$ in terms of boundary proper time $u$. This is the Schwarzian action, which is invariant under $SL_2(\mathbb{R})$ transformations of the trajectory of the `boundary particle' $t=f(u)$, as required by the isometries of the rigid AdS$_2$ spacetime. From this description it is easy to compute the ADM energy of the spacetime~\cite{ Maldacena:2016upp, Engelsoy:2016xyb}, defined as the Noether charge under physical time translations  $u \rightarrow u + \delta u$
\begin{align}
E(u) = - {\bar{\phi}_r \over 8 \pi G_N} \{ f(u), u \}. \label{energygeneral}
\end{align}

Using this diffeormophism, we can construct natural coordinates $y^\pm$ defined by $x^\pm = f(y^\pm)$, in which the metric becomes more complicated,
\begin{equation}\label{eq:ymetric}
	ds^2 = -\frac{4f'(y^+)f'(y^-) dy^+ dy^-}{(f(y^+)-f(y^-))^2},
\end{equation}
but the cutoff is simpler, at constant $\frac{y^+-y^-}{2}=\epsilon$.

The vacuum solutions, with vanishing stress-tensor expectation value\footnote{For the case of conformal matter we are considering, the trace of the stress tensor is a constant determined by the conformal anomaly $ T_{ \ \mu}^\mu  = \frac{c}{24 \pi} R  $, which can be absorbed into the extremal value of the dilaton: $\phi_0 \rightarrow \phi_0^{Ren} = \phi_0 + {c G_N \over 3}$.}, have dilaton profile
\begin{align}\label{eq:BHdilaton}
\phi = 2 \bar{\phi}_r { 1 - {(\pi  T_0)^2} x^+ x^-  \over x^+ - x^-}
\end{align}
up to gauge transformations, which represents an eternal black hole with two asymptotic boundaries and temperature $T_0$ \cite{Almheiri:2014cka}.
The associated reparameterization is
\begin{equation} \label{ebh}
	f(u) = \frac{1}{\pi T_0} \tanh(\pi T_0 u),
\end{equation}
and using this to transform to $y$ coordinates (which cover the exterior of the black hole) the metric and dilaton take the manifestly static form
\begin{align}
ds^2 = {-4 dy^+ dy^- \over {1 \over (\pi  T_0)^2}\sinh^2\left[ \pi  T_0(  y^+ - y^- )\right]}, \ \ \phi = 2 \bar{\phi}_r {\pi  T_0 } \coth \left[  {\pi  T_0 }(  y^+ - y^- )\right].
\end{align}
The boundary particle meets the AdS$_2$ boundary at two locations:
\begin{subequations}
\begin{align}
&\mathrm{for\ }\hspace{0.3cm} u \rightarrow  \infty, \ \ \ \ x^+ = t = {1 \over \pi T_0}, \\
&\mathrm{for \ }\hspace{0.3cm}  u \rightarrow  -\infty, \ \ x^- = t = -{1 \over \pi T_0}.
\end{align}
\end{subequations}
Plugging the reparameterization~\eqref{ebh} into the formula for the energy we find
\begin{align}
E(u) = - {\bar{\phi}_r \over 8 \pi G_N} \left\{ {1 \over \pi  T_0} \tanh \left( {\pi  T_0} u  \right), u \right\} = {\pi \bar{\phi}_r \over 4 G_N} {T_0^2} \equiv E_0
\end{align}
which is the expected leading scaling with temperature of the energy of a near extremal black hole\footnote{This follows if we assume that the entropy of a near extremal black hole is analytic near $T = 0$. Using the first law of thermodynamics implies that energy must be quadratic in $T$.}.

\subsection{Evaporation}\label{ssec:evap}
Starting with the static solution above, we will couple the right boundary to a large external heat bath $B$ at zero temperature, thereby extracting the Hawking radiation and evaporating the right side of wormhole. We describe an explicit model for this coupled evolution in section \ref{sec:matter}, here discussing the consequences which are pertinent to the bulk dynamics, namely the resulting energy-momentum transfer into the black hole.

First, an important transient effect occurs when coupling the right boundary to this external system, namely an initial injection of positive energy into the black hole. This is required to satisfy the ANEC along its horizon, and to prevent the wormhole from becoming traversable.  While the presence of this positive energy is required by consistency, its precise value depends on the details of the system and bath and the precise coupling between the two. We will denote this initial positive energy increase as $E_S$, and find a lower bound on its value.

After this initial `shock' of energy, the energy of the black hole begins to be transferred into the bath via the Hawking radiation. In section \ref{sec:matter} we give a very explicit model for coupling, in which we compute the resulting stress-tensor expectation value. This relies on choosing conformally invariant matter and boundary conditions, so that the stress tensor expectation value is determined by the conformal anomaly. For the energy-momentum expectation value, it gives the same result as the analysis of \cite{Engelsoy:2016xyb}, which used a model of perfect absorption of outgoing Hawking quanta at the boundary. The result (derived in \eqref{eq:T--AdS}) is that after the shock, the ingoing stress-tensor expectation value vanishes in the flat metric $-dy^+ dy^-$, which via the conformal anomaly gives a flux of negative energy in the physical metric \eqref{eq:ymetric}:
\begin{equation}\label{eq:T--}
	 \langle T_{x^-x^-}(x^-)\rangle = E_S\, \delta(x^-)-\frac{c}{24\pi}\{y^-,x^-\}\Theta(x^-)
\end{equation}
This is valid after scaling the Poincar\'e coordinates to set $f'(0)=1$, and we recall that $x^-=f(y^-)$. We can alternatively rewrite the last term using the inversion identity for the Schwarzan, $\{y^-,x^-\} = -f'(x^-)^{-2} \{f(x^-),x^-\}$.

As discussed in \cite{Engelsoy:2016xyb}, this result can be used to solve for the energy of the black hole as a function of boundary time. Varying $I_G + I_{CFT}[g]$ with respect to boundary time yields the energy balance equation
\begin{align}
\partial_u E(u) = f'(u)^2  \left(T_{x^- x^-} - T_{x^+ x^+}\right),
\end{align}
equating the change in energy to the ingoing flux minus outgoing flux. Using this and the expression for energy in terms of the Schwarzian \eqref{energygeneral}, for positive times we find the differential equation
\begin{align}
- {\bar{\phi}_r \over 8 \pi G_N} \partial_u \{ f(u), u \} = {c \over 24 \pi} \{ f(u), u \}, \implies \ \{ t(u), u \} \ \propto \ e^{- k u }
\end{align}
where
\begin{align}
k =    \frac{c}{12} \frac{4G_N}{\bar{\phi}_r} \ll 1.
\end{align}
Putting in the initial energy and the perturbation due to turning on the interaction, the energy as a function of time is found to be
\begin{align}
E(u) =  \Theta(- u) E_0 + \Theta(u) E_1 e^{- k u },
\end{align}
where $E_1 \equiv E_0 + E_S$, and $E_S$ is the positive energy due to turning on the coupling between the system and bath. For small positive time, we have a black hole with new temperature $T_1$ satisfying
\begin{align}
E_1 \equiv {\pi \bar{\phi}_r \over 4 G_N} {T_1^2}.
\end{align}
We can put a bound on the magnitude of $E_S$ by requiring that the new event horizon lies outside the original horizon, so the wormhole does not become traversable.  Since there is no interaction between the left and right boundaries, a traversable wormhole would violate boundary causality.  Moreover, in JT gravity traversable wormholes would require violations of the ANEC \cite{Maldacena:2018lmt,Galloway:2018dak} which, in our context, are forbidden by extending the results of \cite{Faulkner:2016mzt} to Killing horizons in curved space as described in that reference. The new horizon is at $x^+=t_\infty$, where $t_\infty= \lim_{u\to\infty} f(u)$ is the Poincar\'e time at which the boundary particle reaches the boundary, so we require $t_\infty<\frac{1}{\pi T_0}$.

For $u<0$, the reparameterization $f$ is given by the black hole solution \eqref{ebh}. For $u>0$, we must solve the differential equation
\begin{align}
\{f(u), u \} = 2 (\pi T_1)^2 e^{- k u},\qquad f(0)=0,\; f'(0)=1,\; f''(0)=0,
\end{align}
where the initial conditions come from matching to the $u<0$ solution at $u=0$. Explicitly, the solution is
\begin{align}
f(u) = {1 \over \pi T_1}{ - K_0\left[  {2 \pi T_1 \over k}  \right] I_0\left[  {2 \pi T_1 \over k} e^{- k u/2}  \right] + I_0\left[  {2 \pi T_1 \over k}  \right] K_0\left[  {2 \pi T_1 \over k} e^{- k u/2}  \right]  \over  K_1\left[  {2 \pi T_1 \over k}   \right] I_0\left[  {2 \pi T_1 \over k} e^{- k u/2}  \right] + I_1\left[  {2 \pi T_1 \over k}   \right] K_0\left[  {2 \pi T_1 \over k} e^{- k u/2}  \right]},
\end{align}
which gives
\begin{equation}\label{eq:finfty}
t_\infty =  \frac{1}{\pi T_1}{  I_0\left[  {2 \pi T_1 \over k}  \right]   \over   I_1\left[  {2 \pi T_1 \over k}   \right] } = {1 \over \pi T_1} + \frac{k}{4 (\pi T_1)^2} +O(k^2),
\end{equation}
where we have expanded for $k\sim G_N \ll 1$. Our causality requirement $t_\infty < \frac{1}{\pi T_0}$ then gives a lower bound on $E_S$, which we solve or at leading order in $k$:
\begin{align}
{( \pi T_1)^2 }= {( \pi T_0)^2} + {4 \pi G_N \over \bar{\phi}_r} E_S  \implies E_S>\frac{c}{24}T_0 + O(k)
\end{align}
An injection of positive energy on the thermal scale is required to maintain boundary causality.

The complicated Bessel function expression for the reparameterization is only really necessary at very late times, $u\sim k^{-1}\log k$, by which time the black hole is so close to extremality that the semiclassical description breaks down, since we no longer have a parametrically large non-extremal entropy. At times of order $k^{-1}$, when the black hole has evaporated an order one fraction of its mass but remains sufficiently far from extremality, we can approximate $f$ by a simpler form. For this, we use the asymptotic formula
\begin{equation}
	\frac{K_n(z)}{\pi I_n(z)} \sim e^{-2z}\left(1+\frac{4n^2-1}{4z}+ O(z^{-2})\right),
\end{equation}
expanding at small $k$ with fixed $uk$. The result is
\begin{equation}\label{eq:LateReparam}
	\log\left(\frac{t_\infty-f(u)}{2t_\infty}\right) \sim -\frac{4\pi T_1}{k} \left(1-e^{-\frac{k}{2}u}\right) + O(k e^{\frac{k}{2}u}),
\end{equation}
where we have included the correction term relevant at late times; a different correction becomes important at early times, when $u$ is of order one.

Differentiating, we find
\begin{equation}
	\frac{1}{t_\infty -t} f'(u) \sim 2\pi T_1 e^{-\frac{k}{2}u} + O(k^2 e^{\frac{k}{2}u}).
\end{equation}
Taking further derivatives, we can verify that this approximation solves the required equation for the Schwarzian to the specified order. Using these estimates along with the inversion formula for the Schwarzian, we can estimate the stress tensor at these times via
\begin{equation}\label{eq:utLate}
	\{u,t\}\sim \frac{1}{2(t_\infty-t)^2} \left(1+O(k^2 e^{-k u})\right),
\end{equation}
where $t=f(u)$.

With a given stress tensor, the constraints and equation of motion for the metric can be solved in terms of an integral of the stress tensor \cite{Almheiri:2014cka}. To the future of the shock $x^->0$, this solution can be written as
\begin{equation}
	\phi = 2\phi_r \frac{1-(\pi T_1)^2x^+x^- +\tfrac{1}{2} k I(x^+,x^-) }{x^+-x^-},
\end{equation}
where $I$ is the integral
\begin{equation}\label{eq:dilatonInt}
	I(x^+,x^-) = \int_0^{x^-} dt (x^+-t)(x^--t)\{u,t\}.
\end{equation}
Here, we have used the form \eqref{eq:T--} to write the ingoing stress tensor in terms of the Schwarzian, and $u=f^{-1}(t)$. To find quantum extremal surfaces, we will be interested in the variation of the dilaton. For this, we can use the integral expressions
\begin{equation}\label{eq:dilVarInt}
	(x^+-x^-)^2\partial_\pm \left(\frac{I}{x^+-x^-}\right) = \mp  \int_0^{x^-} dt (x^\mp-t)^2 \{u,t\}.
\end{equation}

For early times, using the approximation $f(u)\sim \frac{1}{\pi T_1}\tanh(\pi T_1 u)$, we have
\begin{equation}\label{eq:utEarly}
	\{u,t\}\sim \frac{2(\pi T_1)^2}{(1-(\pi T_1 t)^2)^2},
\end{equation}
which can be used, along with the integral expressions to obtain simple explicit expressions for $I$ and derivatives of the dilaton.

For later times, we can use \eqref{eq:utLate}, which can be seen to match with \eqref{eq:utEarly} at intermediate times; at later times, it will prove vitally important that the double pole is shifted to $t_\infty$, which is corrected from $\frac{1}{\pi T_1}$ by a power series \eqref{eq:finfty} in $k$. We will follow through these calculations as required in section \ref{sec:QES}.


%

\section{The matter sector}\label{sec:matter}

In our model, the matter sector is independent of the dilaton, coupling to it only through the constraints. We can therefore treat it as a quantum field theory on a fixed AdS$_2$ background. In this section we describe our model for the matter and its coupling to the auxiliary system collecting the Hawking radiation. We then compute the quantities relevant for our purposes, namely the stress tensor expectation value, which determines the dynamics of the dilaton, and the entropies of subsystems.

Explicit calculations are made possible by choosing a conformally invariant matter theory, with conformally invariant boundary conditions at the asymptotics of AdS$_2$ before turning on the coupling to the bath. For example, we could choose free massless fields with reflecting boundary conditions.

We will use Poincar\'e coordinates for AdS$_2$:
\begin{equation}\label{eq:xMetric}
	ds^2 = \frac{-dt^2+dz^2}{z^2}= -\frac{4dx^+dx^-}{(x^+-x^-)^2} = \frac{4dxd\bar{x}}{(x+\bar{x})^2},\qquad \begin{cases}x^+=t+z=\bar{x} \\ x^-=t-z=-x. \end{cases}
\end{equation}
The coordinates $x,\bar{x}$ are useful for describing the preparation of our state by a path integral in a Euclidean spacetime with Euclidean time $\tau=it$, so $x=z+i\tau$ and $\bar{x}$ is its complex conjugate. After conformal transformation, the Hartle-Hawking state on AdS$_2$ is given by the vacuum on the half-line $z>0$, where we choose conformally invariant boundary conditions at $z=0$.

\subsection{Coupling to the bath}

Starting with the matter in this state at time $t=0$, we want to allow the black hole to evaporate via a coupling to an auxiliary system which acts as a bath to collect the Hawking radiation. Here, we simply choose the bath to be another half-line supporting the same CFT as the bulk matter theory, also initially in the vacuum with the same conformally invariant boundary condition. At $t=0$, we remove the boundary between AdS$_2$ and the heat bath, allowing matter to move freely between the two.

However, we must be slightly careful when we couple the systems, because we want to match the time evolution in the bath with the physical time evolution of the gravitational system. The physical time $u$ does not match the Poincar\'e time $t$; rather they are related by a diffeomorphism $t=f(u)$, where may choose $f(0)=0$. We must couple the bath at time $u$ to the same physical time at the boundary of AdS$_2$, which corresponds to Poincar\'e time $t=f(u)$.

\begin{figure}
\centering
\includegraphics[width=.46\textwidth]{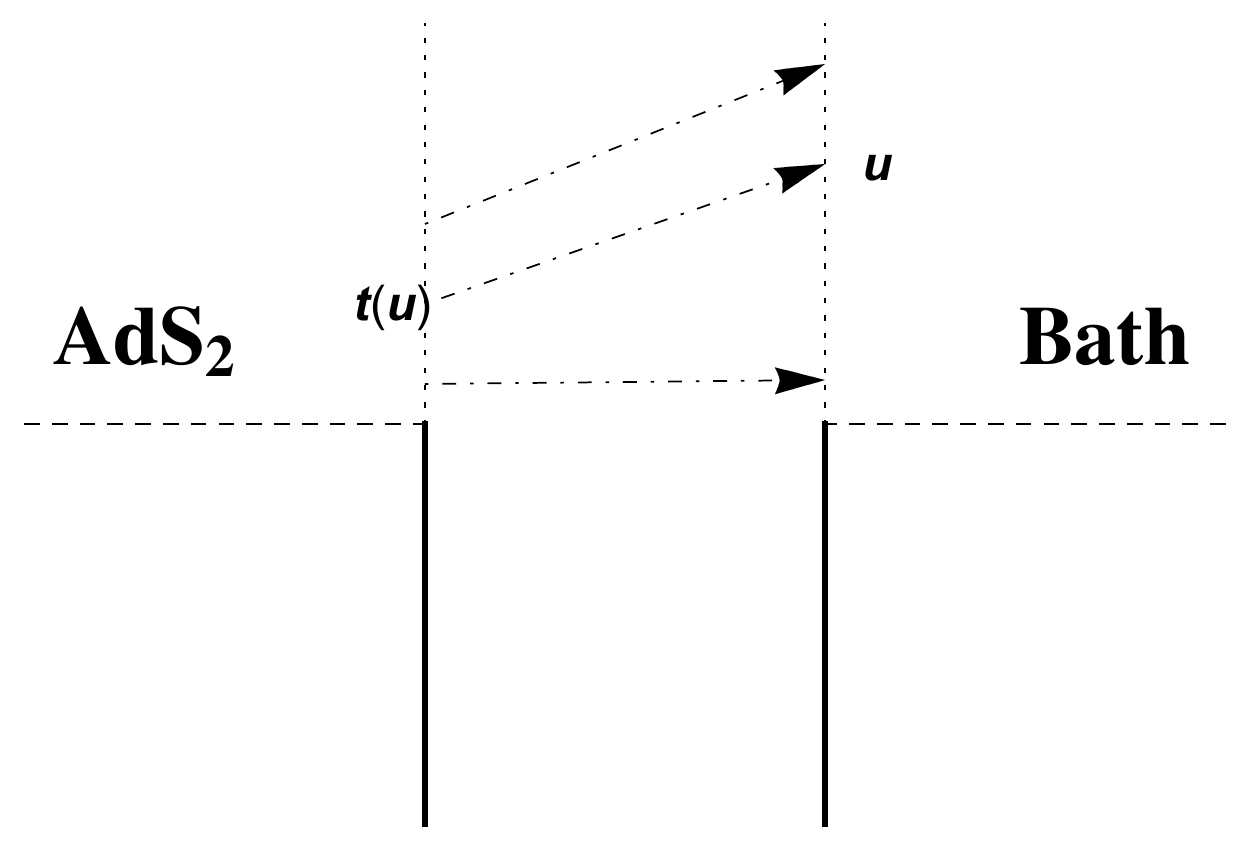} \qquad
\includegraphics[width=.46\textwidth]{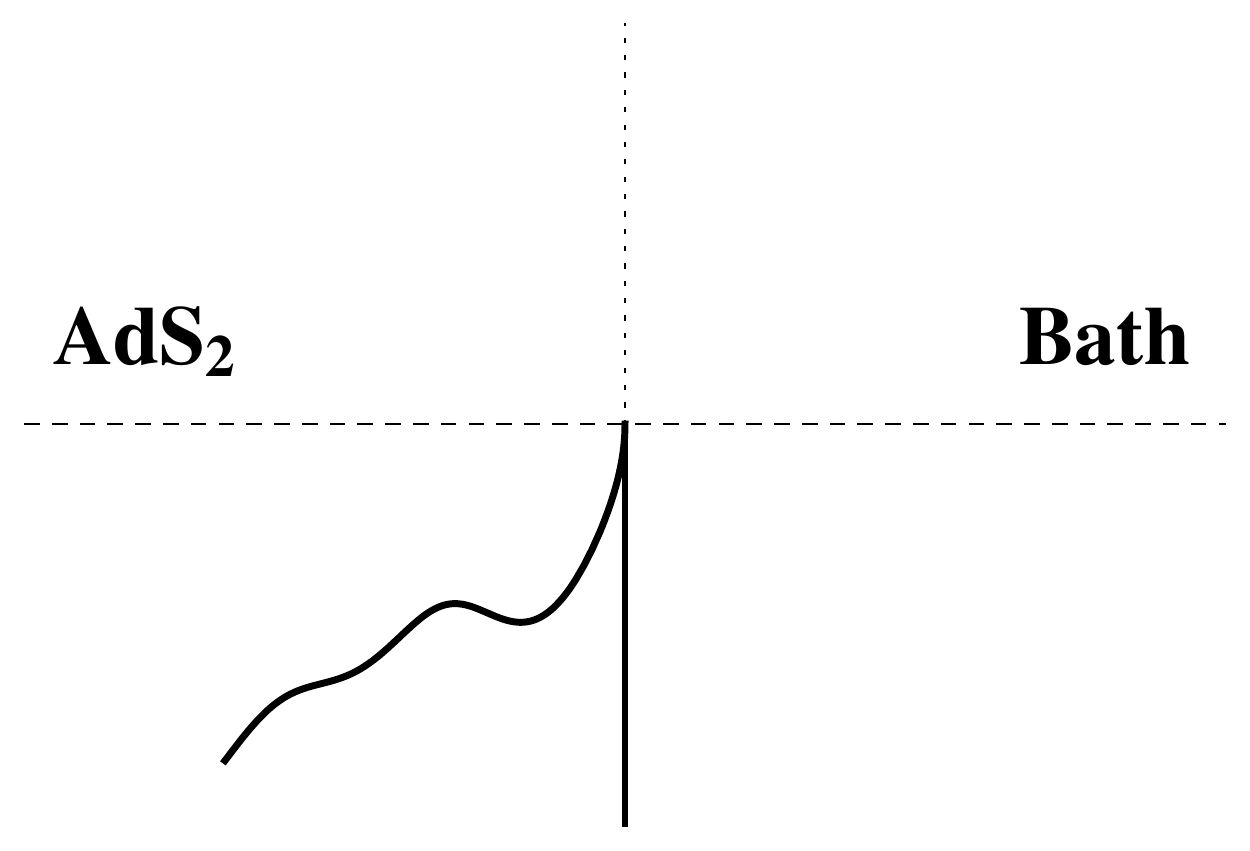}
	\caption{Left: we prepare the state at $t=0$ (dashed line) in both AdS$_2$ and bath in the half-line vacuum, given by the Euclidean path integral in the lower half, but for subsequent time evolution must identify boundary times with the function $u=f(t)$. Right: after making a diffeomorphism in the AdS half, the state is prepared by a path integral with deformed boundary, but time evolution of the coupled system is given simply by the Hamiltonian of the CFT on the line. The lower and upper parts of the right diagram represent the Euclidean and Lorentzian pieces of the path integral in the $y$ coordinates.\label{fig:AdSBath}}
\end{figure}

We can simplify the time evolution by using local conformal symmetry. Namely, we change to coordinates $y,\bar{y}$ such that $x=f(y)$, and $\bar{x}=f(\bar{y})$ on AdS$_2$, in which the metric is more complicated but the boundary coordinate time corresponds to physical time $u$:
\begin{equation}\label{eq:yMetric}
	ds^2  = \frac{4f'(y)f'(\bar{y}) dyd\bar{y}}{(f(y)+f(\bar{y}))^2} = \Omega_y^{-2} dyd\bar{y}
\end{equation}
For the AdS boundary to lie at $y+\bar{y}=0$, and for $\frac{\bar{y}-y}{2}$ to correspond to the physical time $u$ there, we define $f$ for $y<0$ by extending it as an odd function, $f(-y)=-f(y)$. We should regard this as preparing a time-reversal invariant state at $t=0$, which we can evolve symmetrically in either direction with the coupled Hamiltonian; this is different from the physical time evolution, which has a decoupled Hamiltonian and different function $f$ in the past.

We can now make a Weyl transformation using $\Omega_y$ to the flat metric $dyd\bar{y}$, compute in that metric, and transform back at the end. We take the bath to live in the left half-line $y+\bar{y}<0$, with the state prepared by Euclidean path integral on the left half-plane $\Re y<0$. The combined evolution of this coupled system and bath is then implemented by the usual CFT Hamiltonian on the line, and identifies times in the desired way. The price we pay is that the state on the right half-line $y>0$ is a complicated Virasoro descendant of the half-line vacuum, which we can think of as prepared by a path integral in the Euclidean section with a boundary of some complicated shape, as illustrated in figure \ref{fig:AdSBath}.

We note here that the $y$ coordinate will not cover the entire Poincar\'e patch, but only a Rindler patch $x,\bar{x}<f(\infty)$, since $f(y)$ remains finite as $y\to\infty$. The Euclidean path integral preparing the state on the half-line $y\in \RR_+$ will have the topology of a cylinder. A simple example is for $f(u)=\frac{1}{\pi T}\tanh(\pi T u)$, which identifies $u$ with Rindler time; the preparation of the state on AdS$_2$ is then by Euclidean path integral on a half-cylinder, with $y$ periodically identified in imaginary time with period $T^{-1}$, giving a thermal state of matter fields as expected.

\subsection{Mapping to the half-plane}

We now have a description of our initial state on the coupled system and bath, which is characterised by a few simple properties. It is time-reflection symmetric, and a descendant of the vacuum state on the half-line. This follows because it is prepared by the Euclidean path integral on the right of figure \ref{fig:AdSBath}, over a simply connected space with a single boundary. Consequently, there is a diffeomorphism taking the initial Cauchy surface (including both the $t=0$ slice of AdS$_2$ and the bath) to the half-line parameterised by a coordinate $w\in [0,\infty)$, which can be used to map our state to the half-line vacuum. Specifically, our state becomes the vacuum in the half Minkowski space $w+\bar{w}>0$ with lightcone coordinates $w,\bar{w}$ (where we apply the same diffeomorphism to define left- and right-moving coordinates), after applying a Weyl transformation so that the metric becomes $dw d\bar{w}$. Calculation of correlation functions and entropies is then reduced to a calculation in the half-space vacuum, along with a Weyl transformation to the physical metric $ds^2 = \Omega_w^{-2} dwd\bar{w}$.

To identify the diffeomorphism to the new $w$ coordinate, we can use the one-point function of the stress tensor, noting that it vanishes in the half-line vacuum, so $\langle T_{ww}(w)\rangle=0$ (the subscripts denoting the metric we are working in as well as the coordinates, as is conventional in two-dimensional CFT). In AdS$_2$ at $t=0$, the one-point function is zero in the physical metric, and also in the metric $dxd\bar{x}$ since the Weyl anomaly between AdS$_2$ and flat space in Poincar\'e coordinates vanishes; in the bath, the one-point function is zero in the metric $dyd\bar{y}$:
\begin{equation}
	\langle T_{xx}(x) \rangle = 0 \quad (x>0),\qquad \langle T_{yy}(y) \rangle = 0\quad (y<0)
\end{equation}
Now, under most diffeomorphisms the stress tensor picks up an anomaly from the associated Weyl transformation:
\begin{equation}\label{eq:Schwwx}
	\left(\frac{dw}{dx}\right)^2 \langle T_{ww} \rangle = \langle T_{xx}\rangle +\frac{c}{24\pi}\{w,x\}
\end{equation}
Note that we use a standard normalization of the stress tensor, which differs from the common convention in the two dimensional CFT literature by a factor of $-2\pi$. The only diffeomorphisms for which the anomaly is absent, and hence which preserve the vanishing of the stress tensor one-point function, are the M\"obius maps. We therefore find that for the part of the initial data slice in AdS$_2$, $w$ is a M\"obius map of $x=f(y)$, and for the part in the bath, $w$ is a M\"obius map of $y=f^{-1}(x)$. We can now choose the AdS$_2$ region $x>0$ to map to $w\in (0,w_0)$ and the bath region $y<0$ to map to $(w_0,\infty)$, from which we can write the full diffeomorphism as
\begin{equation}
	w(x) = \begin{cases}
		\frac{w_0^2}{w_0+x} & x>0 \\
		w_0 +f^{-1}(-x) & x<0
	\end{cases}
\end{equation}
for some $w_0>0$, where we have scaled the Poincar\'e coordinates $x,\bar{x}$ to set $f'(0)=1$. In writing this, we have required that $w$ and its first derivative are continuous at $x=y=0$, which ensures that correlation functions of primary operators are continuous and that the stress tensor one-point function is physically sensible, as we will see in a moment. We have also made use of the symmetry under rescaling $w$ to fix a free coefficient, and the extension of $f$ to negative values as an odd function.

This map suffices for the piece of AdS$_2$ bounded by the Cauchy horizons at $w$ or $\bar{w}\to \infty$ ($x$ or $\bar{x} \to t_\infty :=\lim_{u\to\infty}f(u)$), and by the Poincar\'e horizons $w=0$ ($x\to\infty$) and $\bar{w}=0$ ($\bar{x}\to\infty$). The map to $w$ coordinates can be straightforwardly extended past the Poincar\'e horizons by using coordinates on AdS$_2$ that cover a larger patch.

\begin{figure}
	\centering
	\includegraphics[width=.3\textwidth]{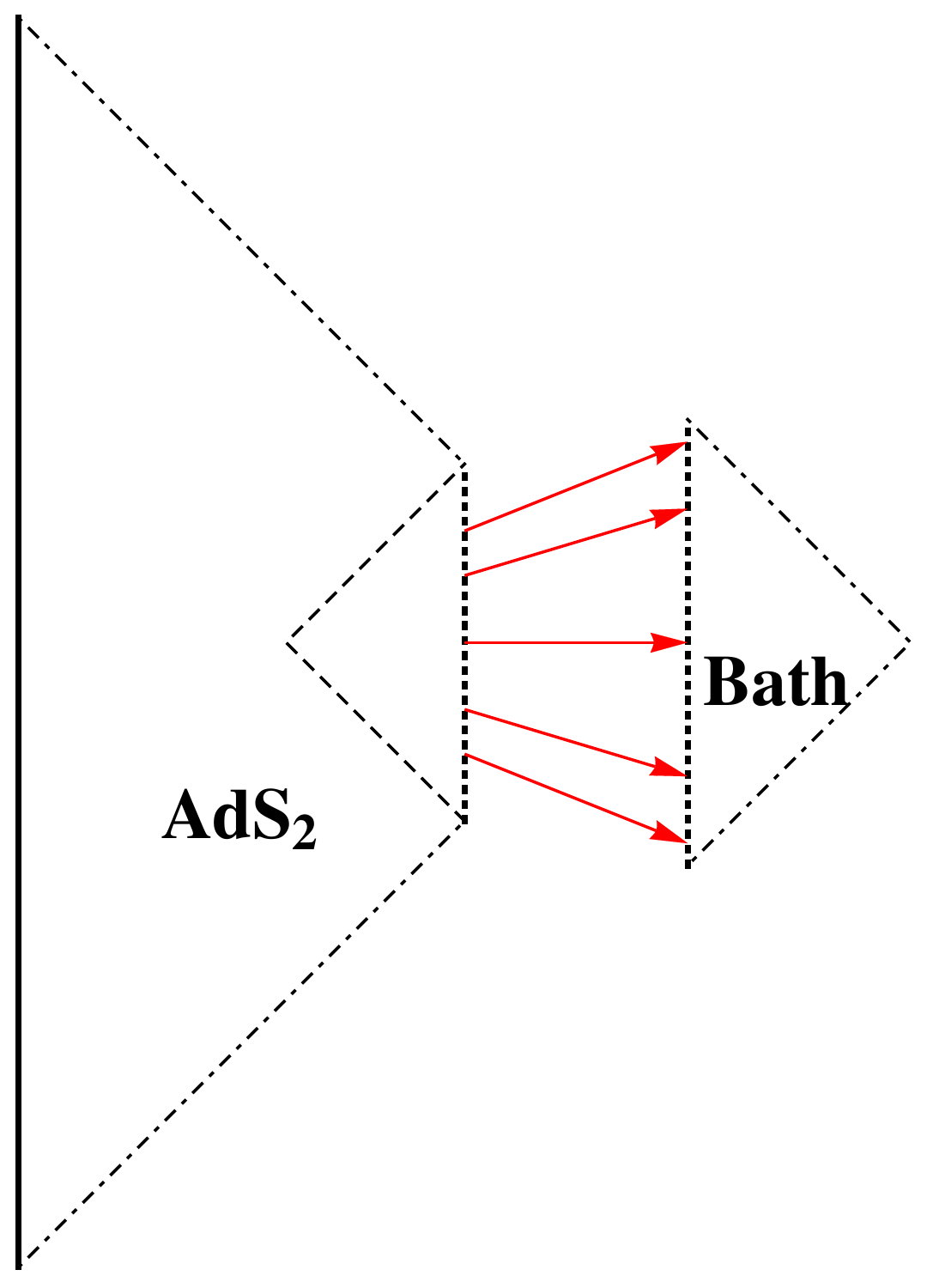}
	 \raisebox{90pt}{$\quad\longrightarrow\quad$}
	\includegraphics[width=.25\textwidth]{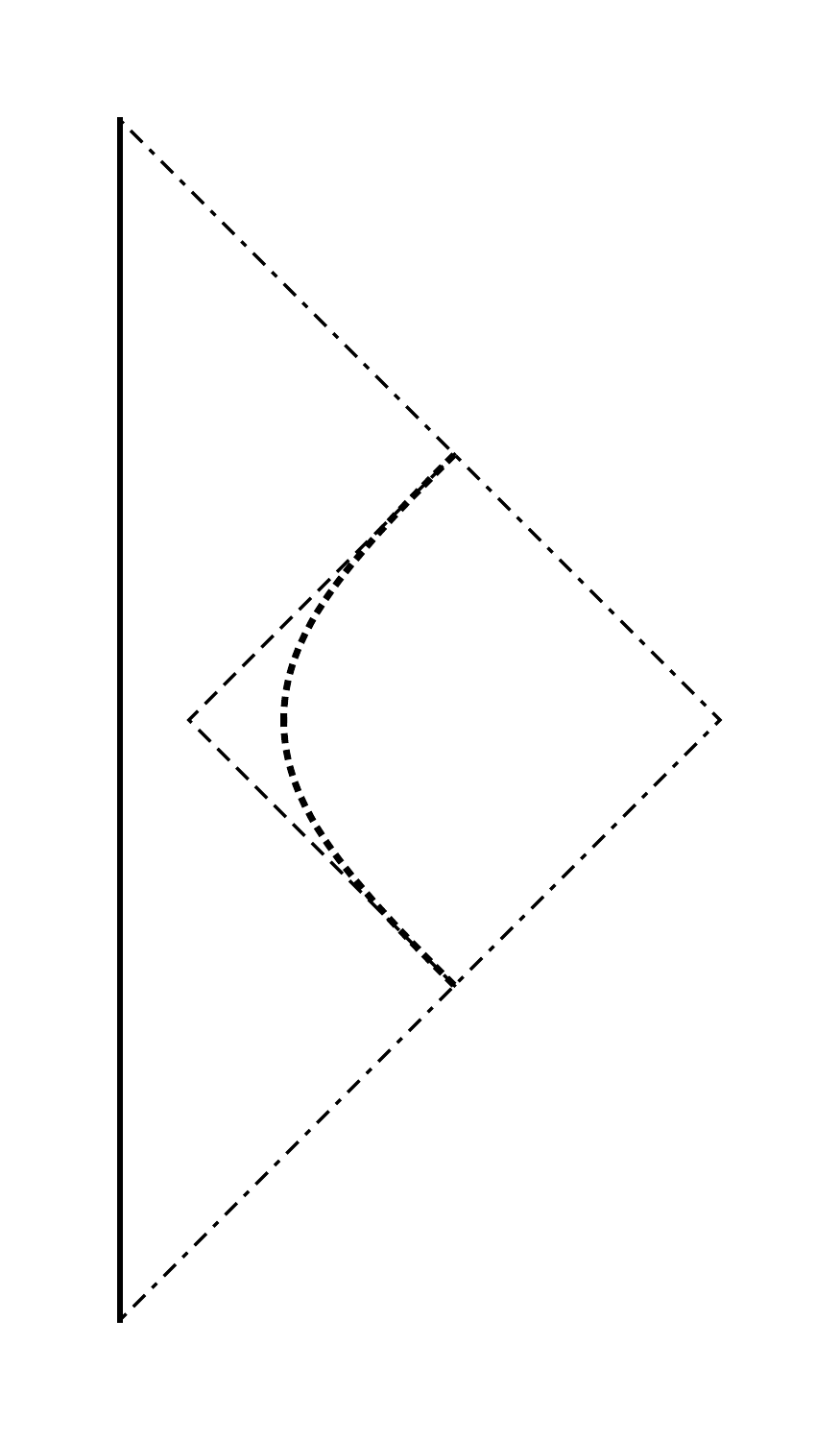}
	\caption{Penrose diagrams of the identification for the time-symmetric coupled system. Only the future-half of the diagram is relevant to the physical system in which the coupling is switched on only at $t=0$. At $t=0$ the bath is prepared in the Minkowski half-line vacuum.  Dash-dotted lines denote future and past null infinity in the relevant Minkowski space. The remaining bath boundary (dotted) is identified with the right boundary of AdS$_2$ using the diffeomorphism $f$. We describe the state of the matter fields in the patch of AdS$_2$ shown, bounded by the dash-dotted lines denoting future and past Cauchy horizons. The dashed lines denote the event horizons of the right boundary. The resulting state is the half-line vacuum in the Minkowski half-space with auxiliary metric $dwd\bar{w}$, for which the Penrose diagram is shown on the right. In the physically relevant limit $w_0\to 0$, the worldline of the joined boundaries is pushed to the left in the $w$ coordinates, becoming nearly null.}
\end{figure}

We can now use this map and the anomaly \eqref{eq:Schwwx} to compute the stress tensor one-point function in the $x$ and $y$ frames (recalling that we have set $f'(0)=1$):
\begin{align}
	\langle T_{xx}(x) \rangle &= -\frac{c}{24\pi} \{w,x\} = E_S \delta(x) -\frac{c}{24\pi} 	\Theta(-x) \{y,x\} \label{eq:Txx} \\
	\langle T_{yy}(y) \rangle &= -\frac{c}{24\pi} \{w,y\} = E_S \delta(y) +\frac{c}{24\pi} 	\Theta(y) \{f(y),y\}  \\
	&E_S =  \frac{c}{24\pi}\left(\frac{2}{w_0}-f''(0)\right)
\end{align}
The $\delta$-function contributions arise from the discontinuous second derivative of $w$, and at this point we are free to tune its coefficient $E_S$ by choice of $w_0$.

However, if we use this result for finite $E_S$, the answers we will find are not compatible with the physics we are trying to capture. In particular, this will give nonzero connected two-point functions for an operator in AdS$_2$ and an operator in the bath, both spacelike separated from the point at which the coupling is turned on, though the states should be uncorrelated at such points. The resolution is that we must take $E_S\to\infty$. This is to be expected since there is of no finite energy state in a quantum field theory that is uncorrelated at finite separation. Without using an explicitly regulated model we must accept either acausal correlations or infinite energy. For example, using a lattice regulated theory where we join a pair of spin chains at time $t=0$, energy will be introduced on the scale of the cutoff, $E_S\sim \varepsilon^{-1}$, and typically there will be faster than light propagation (since the relativistic theory only emerges in the infrared) that is negligible only on scales far longer than the cutoff. In the end, we will take $E_S$ as some intermediate scale between AdS and Planck, the former to avoid dependence on details of the regularization, and the latter for validity of effective field theory, which we presume has a subPlanckian cutoff.

Taking $E_S\to\infty$ ($w_0\to 0$),  we find a simple limiting map to the upper half-plane:
\begin{equation}
	w(x) \sim  \begin{cases}
		 \left(\frac{12\pi}{c}E_S\right)^{-2}\frac{1}{x} & x>0 \\
		 f^{-1}(-x) & x<0
	\end{cases}
\end{equation}

We have precisely the same map for right-moving coordinates $\bar{w}(\bar{x})$.

This approach gives an alternative route to previous results on local quantum quenches  \cite{Calabrese:2007mtj, Calabrese:2009qy, Asplund:2013zba}. These analyses use an explicit regulating prescription, by offsetting the removal of the boundary slightly in Euclidean time. The results are equivalent after identifying $E_S$ with the regulator.

To calculate correlation functions and entropies, we now only need to compute the half-plane correlators in the metric $dwd\bar{w}$, using the maps $w(x),\bar{w}(\bar{x})$, and then transform to the physical metric with the appropriate Weyl factor:
\begin{equation}\label{eq:Weyl}
	ds^2 = \Omega_w^{-2} dwd\bar{w},\qquad \Omega_w = \frac{x+\bar{x}}{2} \sqrt{w'(x)\bar{w}'(\bar{x})} 
\end{equation}

\subsection{Entropy in the half-plane}

Since we have mapped the state of the system to the half-plane, we now need to compute the relevant quantities there, which for us is the entropy of a single interval. We here fix a definition for a renormalized entropy, and then review its computation for a single interval in the half-plane.

A convenient way to compute entropies of intervals in CFT$_2$ is to use the replica trick to compute the R\'enyi entropies $S^{(n)} = -\frac{1}{1-n}\Tr \rho^n$ for integer $n$, as path integrals on $n$ copies of the geometry glued appropriately along the region in question, and take a formal $n\to 1$ limit to recover von Neumann entropy $S=-\Tr \rho\log\rho$. This is particularly powerful for two-dimensional CFTs \cite{Calabrese:2009qy}  because the $n$ copies of the geometry in the theory $\mathcal{C}$ can be described on a single copy of the geometry by correlation functions in an orbifold theory $\mathcal{C}^n/\ZZ_n$, where we take $n$ copies of the original theory and quotient by the symmetry of cyclic permutation\footnote{This is not strictly true in all states, since the orbifold theory correlators may involve an unwanted sum over twisted sectors, but it will hold for all the examples we use.}. In this description, the boundary conditions for the replica manifold are implemented by inserting twist operators $\sigma,\tilde{\sigma}$ at the left and right endpoints of the interval in question:
\begin{equation}
	S^{(n)} = -\frac{1}{n-1} \log \langle\sigma(x_1,\bar{x}_2)\tilde{\sigma}(x_2,\bar{x}_2)\rangle_{\mathcal{C}^n/\ZZ_n}
\end{equation}
We now use the powerful fact that the twist operators are local, primary operators in the orbifold theory, with dimension
\begin{equation}
	\Delta_n = \frac{c}{12}\frac{(n-1)(n+1)}{n} \; .
\end{equation} Since it will be important for us to work on curved manifolds, we emphasize that this property determines correlation functions on a manifold with Weyl rescaled metric $\Omega^{-2}g$ in terms of those with metric $g$ (the expectation values here are normalized by the partition function without operator insertions, so the conformal anomaly cancels):
\begin{equation}\label{eq:sigmaPrim}
	\langle\sigma(x_1,\bar{x}_2)\tilde{\sigma}(x_2,\bar{x}_2)\rangle_{\Omega^{-2}g} = \Omega(x_1,\bar{x_1})^{\Delta_n} \Omega(x_2,\bar{x_2})^{\Delta_n}\langle\sigma(x_1,\bar{x}_2)\tilde{\sigma}(x_2,\bar{x}_2)\rangle_{g}
\end{equation}

The replica manifold is singular at the endpoints of the interval, which leads to a divergent R\'enyi entropy; any local regulator of this divergence can be simply absorbed into the normalization of the twist operators. We can theorefore define renormalized R\'enyi entropies, and hence von Neumann entropies, by choosing a canonical normalization (which implicitly introduces an infrared length scale, for us the AdS scale) where the identity appears in the OPE with coefficient one:
\begin{equation}
	\sigma(x_1)\tilde{\sigma}(x_2)\sim |x_1-x_2|^{-2\Delta_n} \quad\text{ as } x_2\to x_1
\end{equation}
Due to the primary transformation property, this normalization holds independently of the metric, when $|x_1-x_2|$ denotes the proper distance between the operator insertions. In particular, with this normalization, the renormalized entropy of a single interval of length $\ell$ on the line in vacuum is $S=\frac{c}{3}\log \ell$.

Taking the $n\to 1$ limit of \eqref{eq:sigmaPrim}, we find how the von Neumann entropy behaves under Weyl transformations:
\begin{equation}\label{eq:SWeyl}
S_{\Omega^{-2}g} = S_g -\frac{c}{6} \sum_\text{endpoints} \log \Omega
\end{equation}
This can be thought of as arising from a local rescaling of the cutoff scale at each endpoint of the interval.

We can now address the pertinent example of the entropy of a single interval on the half-plane. We begin with the case when the interval contains the boundary, which will be used to compute the entropy in AdS before turning on the coupling, and also for the total entropy of system and bath combined. For this, we insert only one twist operator in the bulk, so we need only write down the most general conformally invariant one-point function on the half-plane:
\begin{equation}
	\langle \sigma(w,\bar{w})\rangle_\text{UHP} = \frac{g_n}{(w+\bar{w})^{\Delta_n}}
\end{equation}
We could think of there being also a twist operator inserted on the boundary, but boundary twist operators are topological: they have zero scaling dimension, and the insertion point can be freely deformed without changing the correlation functions. We note also that the interval of interest could lie on either side of $\sigma$ and give the same result, reproducing the expectation from purity of the total state.

 Taking the $n\to 1$ limit, we find
\begin{equation}\label{eq:UHPS1}
	S = \frac{c}{6}\log(w+\bar{w}) + \log g,
\end{equation}
where $\log g=-\partial_n \log g_n |_{n=1}$ is the Affleck-Ludwig boundary entropy \cite{Affleck_1994}.

Next, we take the interval to have both endpoints in the bulk of the system, away from the boundary. Now, since we are on the half-plane, for two points we can construct a conformally invariant cross ratio (invariant under the $PSL(2,\RR)$ which fixes the boundary on the imaginary axis):
\begin{equation}
	\eta = \frac{(w_1+\bar{w}_1)(w_2+\bar{w}_2)}{(w_1+\bar{w}_2)(w_2+\bar{w}_1)}
\end{equation}
The two-point function of twist-operators does not have a fixed functional form, but contains an undetermined function $G_n(\eta)$, which can depend on the theory and boundary conditions:
\begin{equation}
	\langle\sigma(w_1,\bar{w}_1)\tilde{\sigma}(w_2,\bar{w}_2)\rangle_\text{UHP} = \frac{G_n(\eta)}{\left((w_1-w_2)(\bar{w}_1-\bar{w}_2)\eta\right)^{\Delta_n}}
\end{equation}
In the kinematics relevant for entropy, where the endpoints of the interval remain spacelike separated, the cross-ratio $\eta$ varies between one and zero. Approaching either of these limits, $G_n$ is determined by either a bulk OPE or boundary operator expansion:
\begin{align}
\eta \to 1&,\qquad G_n(\eta)\to 1 \qquad \text{($\sigma\tilde{\sigma}$ OPE limit)}	\\
\eta \to 0&, \qquad G_n(\eta) \to g_n^2 \qquad \text{(boundary limit)}
\end{align}

Taking the $n\to 1$ limit, we find the von Neumann entropy
\begin{equation}\label{eq:UHPS2}
	S = \frac{c}{6}\log\left[(w_1-w_2)(\bar{w}_1-\bar{w}_2)\eta\right] + \log G(\eta),
\end{equation}
where $\log G(\eta) = -\partial_n \log G_n(\eta) |_{n=1}$, satisfying $G(1)=1$, $G(0)=g^2$.

\subsection{Mapping to AdS$_2$}

With all the ingredients in place, it remains only to put them together and compute various quantities of interest in AdS.

First, we compute the expectation value of the stress tensor in AdS$_2$. In the $x$ coordinate in the flat $dxd\bar{x}$ metric this is given by \eqref{eq:Txx}, and there is no anomaly from the Weyl factor to transform to the physical Poincar\'e AdS metric \eqref{eq:xMetric}:
\begin{equation}\label{eq:T--AdS}
	 \langle T_{--}(x^-)\rangle_{\text{AdS}_2} = E_S \delta(x^-)-\frac{c}{24\pi}\{y^-,x^-\}\Theta(x^-),
\end{equation}
where $y^- =f^{-1}(x^-)$. Note that $\{y^-,x^-\}$ will typically be positive, so this represents an injection of negative energy into AdS (in the Hartle-Hawking state on AdS, by $\sl$ invariance the stress tensor expectation value vanishes, aside from the trace which is identically a constant determined by the curvature and anomaly). We can similarly construct $\langle T_{++}\rangle$ by time-reversal invariance, but for the positive times we are interested in it vanishes. This result is used to determine the dynamics of the reparameterization mode $f$ in section \ref{ssec:evap}.

Next, we compute the entropies of an interval in AdS$_2$, along with the bath. This is equal to the entropy of the purifying system, so can be used to verify that the quantum extremal surface of the purifier remains at the bifurcate horizon for all times.

For this, we use the result \eqref{eq:UHPS1}, along with the transformation \eqref{eq:SWeyl}. The result is simplest when the boundary of the interval is spacelike separated from the coupling, $x^+>0$ and $x^-<0$. In that case, we get the same answer as in pure AdS as required by causality, which is a constant by $SL(2,\RR)$ invariance of the spacetime and state:
\begin{equation}\label{eq:constantentropy}
	S = \frac{c}{6} \log 2 + \log g \qquad (x^+>0, \quad x^-<0)
\end{equation}
For a more nontrivial result, we take the endpoint to lie in the future of the shock, $x^+>x^->0$:
\begin{equation}
	S = \frac{c}{6}\log \left(\frac{24\pi E_S}{c}\frac{x^+ y^- \sqrt{f'(y^-)}}{x^+ - x^-}\right) + \log g,\quad x^+>x^-=f(y^-)>0
\end{equation}
An example of this is to compute the total entropy of the bath at physical time $u$ or Poincar\'e time $t=f(u)=\frac{x^++x^-}{2}$, where we cutoff AdS$_2$ at $z=\frac{x^+-x^-}{2} = \epsilon f'(u)$:
\begin{equation}\label{eq:bathS}
	S = \frac{c}{6}\log \left(\frac{12\pi E_S}{\epsilon c}  \frac{u f(u)}{\sqrt{f'(u)}}\right) + \log g
\end{equation}

Finally, we compute the entropy of an interval in AdS$_2$ using \eqref{eq:UHPS2}. There are two main cases of interest, with either one or both endpoints to the future of the shock. We begin with the case with one endpoint to the future, $x_1^{\pm}>0$, and one endpoint to the past, $x_2^+>0,x_2^-<0$. In this case, the cross-ratio is nontrivial,
\begin{equation}
	\eta = \frac{x_1^+(x_2^+-x_2^-)}{x_2^+(x_1^+-x_2^-)}
\end{equation}
and the entropy is
\begin{equation}\label{eq:entropy1}
	S=\frac{c}{6}\log\left[\frac{48\pi E_S}{c}   \frac{ -y_1^- x_1^+x_2^-(x_2^+-x_1^+)\sqrt{f'(y_1^-)}}{x_2^+(x_1^+-x_1^-)(x_1^+-x_2^-)} \right] + \log G(\eta).
\end{equation}
If we take the whole interval to lie to the future of the shock, $x^{(1,2)}_{\pm}>0$, the cross ratio goes to $\eta=1$, and we have
\begin{equation}\label{eq:entropy2}
	S = \frac{c}{6}\log\left[   \frac{4(y_1^--y_2^-)(x_2^+-x_1^+)\sqrt{f'(y_1^-)f'(y_2^-)}}{(x_1^+-x_1^-)(x_2^+-x_2^-)} \right].
\end{equation}
As a consistency check, we note that all entropies are invariant under the residual symmetries of AdS$_2$, namely the $PSL(2,\RR)$ transformations $x\mapsto \frac{ax}{cx+d}$ that fix zero. This means that we act on all coordinates labelled by $x$, and on $f$, but not on $y$.

Finally, take the limit of the above answers where one endpoint goes to the AdS boundary at physical time $u$, Poincar\'e time $t=f(u)$, regulated to lie on the cutoff surface $z=\epsilon f'(u)$:
\begin{equation}\label{eq:Sresults}
S=
	\begin{cases}
		\frac{c}{6}\log\left[\frac{24\pi E_S}{\epsilon c}   \frac{ -u t x^-(x^+-t)}{x^+(t-x^-)\sqrt{f'(u)}} \right] + \log G\left(\frac{t(x^+-x^-)}{x^+(t-x^-)}\right) & x^-\!<0<t<x^+ \\[15pt]
			\frac{c}{6}\log\left[   \frac{2(u-y^-)(x^+-t)}{\epsilon (x^+-x^-)} \sqrt{\frac{f'(y^-)}{f'(u)}} \right] & 0<x^-\!<t<x^+
	\end{cases}
\end{equation}
This will be our main result required for finding quantum extremal surfaces.

\subsection{Interpretation of entropies}

Our formulas for the entropies admit simple quasiparticle interpretations in terms of freely propagating independent left- and right-moving degrees of freedom.

In \eqref{eq:entropy2}, factor $(x^+_{2}-x^+_{1})$ gives the entropy of outgoing modes in the Poincar\'e vacuum. The factor $(y^-_{1}-y^-_{2})$ computes the entropy of modes moving in from the bath, which are in the vacuum associated to the flat metric of the $y$ coordinates. The remaining terms are conformal factors to transform to the physical AdS$_2$ metric, quantifying short-distance correlations. This can be thought of as adapting the cutoff to correspond to a local proper distance.

The case \eqref{eq:entropy1} when the interval straddles the shock allows for a similar interpretation. The outgoing modes are the same as for the previous case of \eqref{eq:entropy2}. The factor $y^-_{1}$ gives the entropy of the infalling state from the bath, starting only at time zero once the coupling to the bath is turned on. For the rest of the interval, the ingoing modes have been reflected off the boundary, so their entropy includes contributions from entanglement with outgoing modes, which is quantified using \eqref{eq:UHPS2}. The constant factor, proportional to $\log E_S$, represents the entropy of the shock itself, which is entangled with the corresponding shock propagating into the bath.

\section{The quantum extremal surfaces}\label{sec:QES}

Now that we have computed the entropy of matter fields and solved the gravitational dynamics, we have all the required ingredients to locate the quantum extremal surface, and hence the entanglement wedge as the black hole evaporates.

\subsection{The generalized entropy}

A quantum extremal surface is defined to extremize the generalized entropy, which is the sum of area in Planck units and the entropy of bulk matter. However, on first sight this is ambiguous, because both ingredients depend on choices of regulator, or the energy scale of the bulk effective field theory we choose to use. The resolution is that the generalized entropy is in fact better defined than its constituents (see~\cite{Bousso:2015mna} for a recent discussion), which can be seen very explicitly in our model.

First, we can define the generalized entropy in terms of renormalized, or infrared quantities:
\begin{equation}
	S_\text{gen} = \frac{\phi_0^{\text{(Ren.)}}+\phi^{\text{(Ren.)}}}{4G_N} + S^{\text{(Ren.)}}
\end{equation}
For example, we can define the renormalized entropy $S^{\text{(Ren.)}}$ as a finite quantity using the scheme of section \ref{sec:matter}. This implicitly involved introducing a finite, infrared scale, in our case the AdS scale.

For an alternative definition, we can use the `bare' quantities, explicitly cutting off the bulk matter theory at some UV scale $\varepsilon$. For our calculations, we treat the gravitational sector classically, but integrating out the matter sector will nonetheless renormalize the gravitational parameters.


 Even though our matter sector is conformal, regularization must introduce a length scale. Explicitly, the CFT partition function on some space with Euler character $\chi$ depends on the typical length scale $L$ (for us, $L$ is the AdS scale):
\begin{equation}
	\mathcal{Z}_\text{CFT}[L] \propto \left(\frac{L}{\varepsilon}\right)^{\frac{c}{6}\chi}
\end{equation}
The functional dependence on $L$ is determined by the trace anomaly, but it necessitates the introduction of the length scale $\varepsilon$. Note that this has the same form as the topological term in the action:
\begin{equation}
	-\frac{\phi_0}{16\pi G_N}\left[\int \mathcal{R} + 2\int_\partial K\right] = -\frac{\phi_0}{4G_N}\chi
\end{equation}
We can therefore absorb this scale dependent normalization of the partition function by an additive renormalization of the dilaton.
\begin{equation}
	-\frac{\phi_0^{\text{(Bare)}}}{4G_N}\chi - \log\mathcal{Z}_\text{CFT} = -\frac{\phi_0^{\text{(Ren.)}}}{4G_N}\chi, \qquad \phi_0^{\text{(Ren.)}} =\phi_0^{\text{(Bare)}} - \frac{2}{3}c\, G_N  \log\varepsilon
\end{equation}
If we use the same cutoff scale for the entropies, the bare and renormalized quantities are related by a similar shift
\begin{equation}
	S^{\text{(Bare)}} = S^{\text{(Ren.)}} - N \frac{c}{6}\log \varepsilon,
\end{equation}
where $N$ is the number of endpoints of an interval in question. Combining these two results, we see explicitly that we can use bare quantities in the definition of the generalized entropy, and find a finite, regulator independent result that matches the one obtained using renormalized, infrared quantities.

\subsection{Early times}\label{sec:earlyQES}

Before we have disturbed the black hole by coupling to an external system, the quantum extremal surface coincides with the classical extremal surface, at the bifurcation point of the original black hole horizon. For a time after the coupling is turned on, the QES does not stray far; we begin by finding its location at these early times.

The bifurcation surface is of course causally disconnected from the process of coupling to the auxiliary system, so the nearby geometry is unaffected, but the quantum extremal surface can nonetheless begin to move as soon as the evaporation process begins. The relevant effect is that the fields near the bifurcation surface are entangled with the first few Hawking quanta to escape; an entangling surface closer to the boundary captures less of this entanglement, and hence has lower entropy. Moreover, this entanglement changes linearly with distance, whereas the classical area term varies quadratically, with the result that the quantum extremal surface moves out towards the boundary in a spacelike direction, of order a Planck distance if evaporation proceeds for a thermal time.

We now show this quantitatively in our model. For a surface located in the $x^-<0$ region before the shock, the area is simply that of the original black hole, while the entropy is computed from the first case in \eqref{eq:Sresults}.
\begin{equation}
\begin{gathered}
	 S_\text{gen} = \frac{\phi}{4G_N} + S, \qquad\qquad  \phi = \phi_0 + 2\bar{\phi}_r\frac{1-(\pi T_0)^2x^+x^-}{x^+-x^-},  \\
 S= 2k\frac{\bar{\phi}_r}{4G_N} \left[ \log\left(\frac{24\pi E_S}{\epsilon c}  \frac{-x^-(x^+ -t) t u}{x^+(t-x^-)\sqrt{f'(u)}}\right)+ \mathcal{G}\left(\frac{t(x^+-x^-)}{x^+(t-x^-)}\right) \right]
\end{gathered}
\end{equation}
We have here defined $\mathcal{G}(\eta) = \frac{6}{c} \log G(\eta)$ for brevity of notation; in particular $\mathcal{G}(1)=0$.

We now simply make the ansatz that the deviation from the horizon is small, taking $x^\pm \mp \frac{1}{\pi T_0}$ of order $k$, and solve for stationarity of $S_\text{gen}$ in the $k\to 0$ limit.
\begin{align}
	\frac{1}{\pi T_0}-x^+ &\sim \frac{k}{(\pi T_0)^2}(\eta-\eta(1-\eta) \mathcal{G}'(\eta))\\
	 x^-+\frac{1}{\pi T_0} &\sim \frac{k}{(\pi T_0)^2} \left(\frac{\eta}{1-\eta}-\eta \mathcal{G}'(\eta)\right) \\
	  \eta &=\frac{2\pi T_0 t}{1+\pi T_0 t}
\end{align}
This solution describes a QES that starts at the bifurcation surface at $t=0$, moves outwards in a spacelike direction towards the boundary, before bending to an almost null path that runs along the horizon, approaching constant outgoing coordinate $x^+$.

When parameterized in terms of the Poincar\'e time $t$ on the boundary, this family of surfaces moves steadily along the horizon, and can be continued far away from the bifurcation surface, beyond the point where our approximation breaks down. However, in terms of the physical time $u$, they end up settling at final location, because as $u\to\infty$, $t$ approaches a finite limit $t_\infty\sim \frac{1}{\pi T_1}<\frac{1}{\pi T_0}$. In the limit where the shock from turning on the coupling adds much more than a thermal unit of energy ($E_S\gg T_0$, or equivalently $T_1-T_0 \gg k$), this location is sufficiently close to the bifurcation surface to be within the regime of our approximation:
\begin{equation}
	1-\eta \gtrsim \frac{T_1-T_0}{T_1+T_0}
\end{equation}
For sufficiently early times, we have $1-\eta \sim e^{-2\pi T_1 u}$, until $\eta$ settles down to its maximum for $u\gtrsim -\frac{1}{T_1}\log\left(\frac{T_1-T_0}{T_1+T_0}\right)$.

We can now evaluate the generalized entropy on the extremal surface, to track the fine-grained entropy of the black hole (this equation valid for times $u\ll k^{-1}$):
\begin{equation}\label{eq:SGenPre}
	S_\text{gen} \sim \frac{\phi_0 + 2\pi T_0
	\bar{\phi}_r}{4 G_N} + \frac{c}{6} \log\left(\frac{24\pi E_S}{\epsilon c} \frac{u\sinh(\pi T_1 u)}{\pi T_1} (1-\eta)\right) + \frac{c}{6} \mathcal{G}(\eta)
\end{equation}
At vey early times, we have a large logarithmic increase in the entropy, due to production of short-distance entanglement with the bath; we should take this seriously only for times larger than the regulator scales $\epsilon$ and $E_S^{-1}$, so the entropy here increases from the equilibrium value. Next, the $\log(1-\eta)$ term takes over, giving a linear decrease in entropy as thermal Hawking radiation escapes into the bath. Eventually, after a scrambling time
\begin{equation}
\label{eq:tSmin}
t_{\text{min} \ S} = \frac{1}{2\pi T_1}\log\left(\frac{T_1}{T_1-T_0}\right) = \frac{\beta_1}{2\pi}\log \left(\frac{E(T_1)}{E_S} \right)
\end{equation}
for $E(T_1)$ the energy of a black hole at temperature $T_1$ and  $T_1  - T_0 \ll T_0$, $\eta$ settles down to its maximum, entropy reaches a local minimum, and then entropy begins to increase linearly. The system begins to access all the extra states opened up by the addition of energy, so we have an increase in thermal entropy purified by the outgoing Hawking radiation, at rate $\frac{c}{6}\pi T_1$. The black hole is now `young', in the sense that its fine grained entropy is less than the thermodynamic entropy at same energy. This linear increase is the first part of the familiar Page curve \cite{Page:1993df}, though for an old black hole after a perturbation rather than for a hole formed recently from collapse.  We note that the time \eqref{eq:tSmin} is precisely of the form anticipated in section \ref{sec:HoloHP} with coefficient $\alpha_ {S}=1$.
\begin{figure}
	\centering
	\includegraphics[width=.5\textwidth]{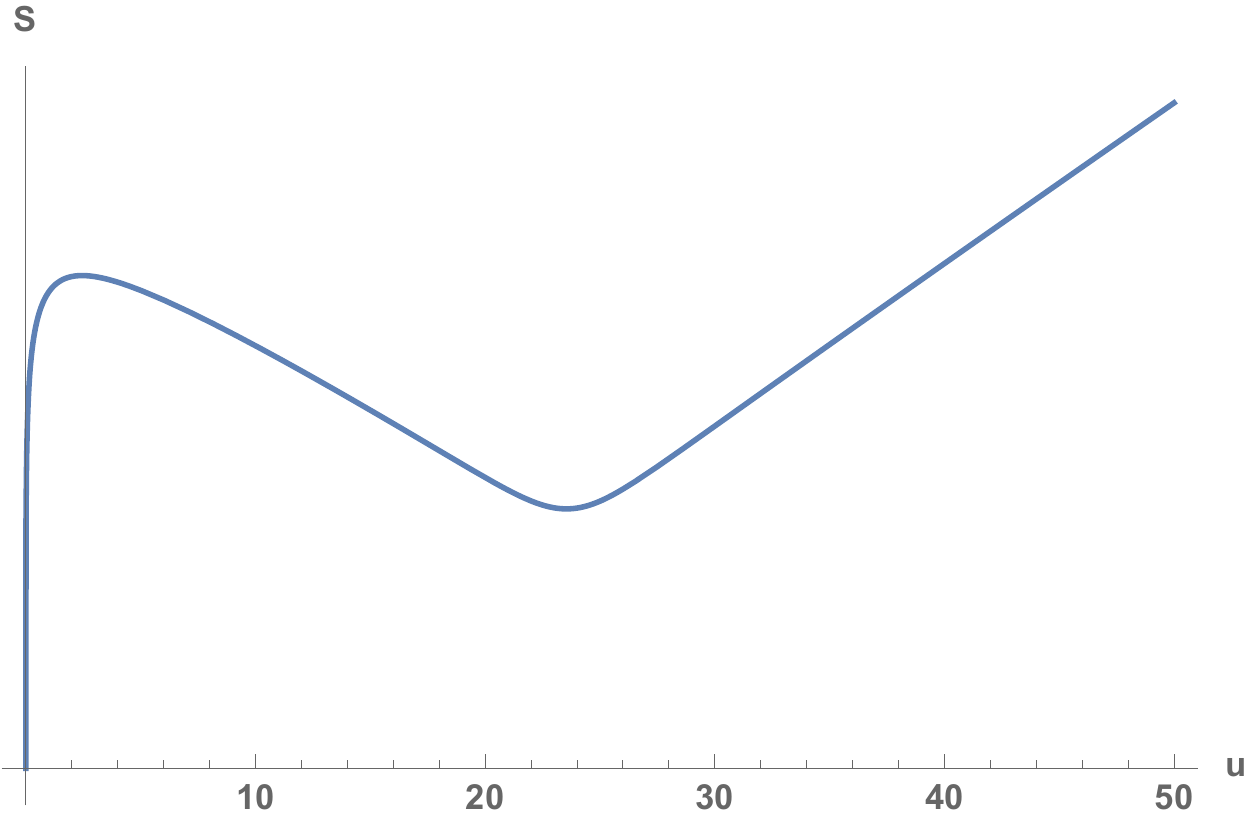}
	\caption{The von Neumann entropy of the black hole, as computed from the generalized entropy on the quantum extremal surface, at early times after coupling to the bath. In this plot, we have $T_0=\frac{1}{2\pi}$ and $\frac{T_1-T_0}{T_0}=10^{-10}$.}
\label{fig:EarlySgen}
\end{figure}

The surfaces described here are the only relevant solutions to the quantum extremization equations for $x^-<0$. There exists another family of surfaces at small negative $x^-$ which extremize $S_\text{gen}$, but these always have much larger generalized entropy.

\subsection{A QES inside the shock}

The next place for a quantum extremal surface to exist is inside the shock itself, at $x^-$ very close to zero. While we do not have a microscopic model for the entropy within the shock itself, these details will not matter, and we can focus on the generalized entropy for small negative or positive $x^-$, just before or after the shock falls in.

Expanding the entropy near the shock, we have the following:
\begin{equation}
	S \sim \frac{c}{6}\log\left[\frac{2u(x^+-t)}{\epsilon x^+\sqrt{f'(u)}}\right] + \frac{c}{6}\begin{cases}
		\log\left(\frac{12\pi E_S}{ c}  \frac{-x^-}{1}\right) &  x^-<0\\
		\left(\frac{1}{x^+}-\frac{1}{u}\right)x^- & x^- >0
	\end{cases}
\end{equation}
We should take the first line seriously only for $x^- \gg E_S^{-1}$ (which can be very close to the shock because we are taking $E_S\gg 1$). For an interval that straddles the shock, there is a large, $\log E_S$ contribution to the entropy, from entanglement between the shock falling into the black hole and its counterpart propagating into the auxiliary system. The entropy drops off rapidly as we approach $x^-=0$, since the initial product state between system and bath carries no short-distance entanglement.

To look for a quantum extremal surface, we first extremize the generalized entropy in the $\partial^+$ direction, using the dilaton $\phi-\phi_0 = \frac{2\bar{\phi}_r}{x^+}$:
\begin{equation}
	\partial_+ S_\text{gen} =0 \implies x^+ = \frac{t}{1-k t}
\end{equation}
 Since the dilaton is decreasing as a function of $x^+$, we need to balance this against a rapid increase from the entropy. This comes from the lightcone singularity as $x^+$ approaches $t$, a phenomenon we will revisit later. At this location, the generalized entropy is maximal for variations parallel to the shock.

 For a quantum extremal surface to exist on the shock, we will also require that the generalized entropy is locally minimal under variations in the outgoing, $x^-$ direction:
 \begin{equation}
 	\frac{4G_N}{\bar{\phi}_r}\partial_- S_\text{gen} \sim  \begin{cases}2\frac{1}{(x^+)^2}-2(\pi T_0)^2+
 		\frac{2k}{x^-} & x^-<0 \\ 2\frac{1}{(x^+)^2}-2(\pi T_1)^2+ \frac{2k}{x^+}-\frac{2k}{u} & x^->0
 	\end{cases}
 \end{equation}
 It appears that the singularity in the entropy will cause this to be decreasing just before the shock, but generically this will only be large enough to overcome the large dilaton increase within the microscopic scale of the shock $E_S^{-1}$, so the result should not be trusted. However, this is no longer true very close to the horizon, for $\frac{1}{\pi T_0}-x^+  \ll \frac{kE_S}{T_0^3}$, in which case we can have a quantum extremal surface on the condition that the entropy begins to increase again for $x^->0$. This is satisfied as long as
 \begin{equation}
 	x^+-\frac{1}{\pi T_1} \lesssim \frac{k}{2(\pi T_1)^2} \implies t-\frac{1}{\pi T_1} \lesssim -\frac{k}{2(\pi T_1)^2}
 \end{equation}
Translating this into the boundary proper time, using the expression \eqref{eq:LateReparam} for $t$ expanded to first order in small $ku$, and the expression \eqref{eq:finfty} for $t_\infty$, we find that there is a candidate quantum extremal surface supported on the shock for times
\begin{equation}
	u< \frac{1}{2\pi T_1} \log\left(\frac{8\pi T_1}{3k}\right).
\end{equation}

The generalised entropy for these surfaces is larger at any given boundary time than the corresponding entropy \eqref{eq:SGenPre} for the extremal surface we found in the previous section. This is simply because the area has been increased by the energy from the shock, and the additional entropy is far too small to make up for it. These surfaces can never have minimal $S_\text{gen}$, so are not relevant for determining the entanglement wedge.

\subsection{Soon after the shock}\label{ssec:soon}

We now look for quantum extremal surfaces in the region $x^->0$, to the future of the coupling to the bath, and the resulting shock of energy. At this point, if one thinks of a quantum extremal surface as a small perturbation to a classical extremal surface, it may seem that none should exist in this regime. In particular, for any $x^->0$, the area is never close to being stationary under variations in the $+$ direction.

Nonetheless,  the bulk entropy can compete with the area to introduce an extremum of the generalized entropy which is far from any classical extremal surface. The necessary enhancement of the variation of entropy comes from the large amount of entanglement in the bulk fields at short distances, in particular in the outgoing Hawking radiation. Consider the generalized entropy of a surface at some fixed $x^-$ as we decrease $x^+$, moving the surface out towards the boundary in a past null direction. The area steadily increases all the way, but the entropy at some fixed boundary time decreases, from the loss of entanglement with the previous Hawking radiation. When the surface approaches null separation from the boundary time of interest, the rate of decrease of entropy is sufficient to overcome the increase in area, so the generalized entropy has a maximum in this direction. This will typically happen close to the apparent horizon, in which case the variation of area along the horizon is in any case small. It is then unsurprising that the area can also be extremized in the direction parallel to the horizon, leading to our quantum extremal surface.

To see this more quantitatively, take first the case when $x^-$ is of order one. At this early time, for many purposes we can ignore the backreaction resulting from the slow leakage of energy into the bath, so can take the dilaton profile of the static black hole solution with temperature $T_1$ (including the effect of the shock). For extremizing in the $x^+$ direction, the only relevant piece of the bulk entropy is proportional to $\log(x^+-t)$, quantifying the entanglement of outgoing modes. Keeping only these terms, we find
\begin{align}
	\frac{4G_N}{\phi_r}&\partial_+ S_\text{gen} \sim -2 \frac{1-(\pi T_1 x^-)^2}{(x^+-x^-)^2} +\frac{2k}{x^+-t} \nonumber \\
	&\implies x^+ -t \sim \frac{k}{(\pi T_1)^2}\frac{1-\pi T_1 x^-}{1+\pi T_1 x^-}, \label{eq:earlyPlusVar}
\end{align}
where in the last equation we have also approximated $t,x^+\sim \frac{1}{\pi T_1}$, which is required for the quantum extremal surface to exist in the $x^- >0$ region.

Now it remains only to extremize in the $x^-$ direction, along the horizon. Close to the horizon, the variation of the background dilaton is suppressed (proportional to $x^+-\frac{1}{\pi T_1}$), so we must also include the backreaction:
\begin{equation}
\begin{aligned}
	(x^+-x^-)^2 \partial_- \left(\frac{I}{x^+-x^-}\right) &\sim \int_0^{x^-} dt(x^+-t)^2 \left\{ \frac{\tanh ^{-1}(\pi  t T)}{\pi  T},t \right\} \\
	& \sim \frac{2x^-}{1+\pi T_1 x^-}
\end{aligned}
\end{equation}
Combining this with the unbackreacted term, where we use \eqref{eq:earlyPlusVar} to determine the deviation of $x^+$ from the horizon, we find
\begin{equation}
	\partial_- \phi \sim \bar{\phi}_r \left(\frac{(2\pi T_1)^2(1 - \pi T_1 x^+)}{(1-\pi T_1 x^-)^2}+ \frac{2 (\pi T_1)^2kx^-}{(1+\pi T_1 x^-)(1-\pi T_1 x^-)^2}\right)
\end{equation}
Finally, we need the variation of the entropy. We use the early time, unbackreacted solution of the reparameterization, $f(u)\sim \frac{1}{\pi T_1} \tanh(\pi T_1 u)$, so in particular we have $f'(y^-)\sim 1-(\pi T_1 x^-)^2$. For times $u\gg 1$, the variation of the $\log(u-y^-)$ term is negligible.
\begin{equation}
	\partial_- S = \frac{c}{6} \frac{\pi T_1}{1-(\pi T_1 x^-)^2} 
\end{equation}
Putting this together we find
\begin{equation}
	\frac{4G_N}{\bar{\phi}_r}\partial_- S_\text{gen} \sim  \frac{(2\pi T_1)^2}{(1-\pi T_1 x^-)^2}\left[(1-\pi T_1 x^+)+\frac{k}{2\pi T_1}\frac{1}{1+\pi T_1 x^- }\right],
\end{equation}
which gives us our condition for quantum extremality in the $-$ direction:
\begin{equation}
	\partial_- S_\text{gen}=0 \implies x^+ -\frac{1}{\pi T_1} \sim \frac{k}{2(\pi T_1)^2} \frac{1}{\pi T_1 x^-+1}
\end{equation}

This locates our quantum extremal surface, which we now write in terms of proper time $u$. In the relevant regime, with $ke^{2\pi T_1 u}$ of order one, this is related to Poincar\'e time by $t-\frac{1}{\pi T_1}\sim \frac{k}{(2\pi T_1)^2}- \frac{2}{\pi T_1} e^{-2\pi T_1 u}$ (using \eqref{eq:LateReparam}, and requiring the expansion \eqref{eq:finfty} of $t_\infty$).
\begin{align}
	x^- &\sim \frac{1}{\pi T_1} \frac{3k e^{2\pi T_1 u}-8\pi T_1}{3k e^{2\pi T_1 u}+8\pi T_1} \\
	t_\infty - x^+  &\sim  -\frac{2}{3}t_\infty e^{-2\pi T_1 u}
\end{align}
For this to be valid, we require that the extremum lies to the future of the shock in the $x^->0$ region, so $u>\frac{1}{2\pi T_1} \log\left(\frac{8\pi T_1}{3k}\right)$. Before this time, this family of surfaces joins continuously onto those of the previous section, which live somewhere inside the shock.

\subsection{Later times}\label{ssec:later}

The quantum extremal surfaces found so far have relied on approximations to the dilaton profile which are valid only at early times, and our result holds only when $e^{-2\pi T u} \gg k^2$. Going to higher orders in the small parameter $k$ does not help much: it would extend validity to times when $e^{-2\pi T u}$ scales as a higher power in $k$, which accesses only a few scrambling times. To go to times of order $k^{-1}$, when a significant fraction of the black hole has evaporated, we require a different approach to treat the dilaton. The idea is to determine the dilaton by working backwards from very late times, and using the fact that it vanishes on the boundary at the Poincar\'e time $t=t_\infty$, corresponding to $u\to \infty$ in proper time. This prevents the exponential growth of errors in the approximations we make.

 We continue the same intuition that the lightcone singularity balances the area in the ingoing direction, making the ansatz $ t_\infty -x^+ \ll t_\infty -x^-$ to put the surface close to the horizon. Begin with the variation of the dilaton in the plus direction, using the integral expression \eqref{eq:dilVarInt} for the effect of backreaction:
\begin{equation}
	(x^+-x^-)^2 \partial_+ \phi = \bar{\phi}_r \left[2(\pi T_1 x^-)^2-2-k \int_0^{x^-} dt (x^--t)^2 \{u,t\} \right]
\end{equation}
The right hand side is a function of $x^-$ only, which must be negative for any $x^-<t_\infty$, since $\phi$ goes to infinity at the boundary. At $x^-=t_\infty$, it vanishes, so in that limit the integral must give
\begin{equation}
	I_\infty = \int_0^{t_\infty} dt (t_\infty-t)^2 \{u,t\} = \frac{2}{k}((\pi T_1 t_\infty)^2-1).
\end{equation}

We can now work backwards from this result to estimate the integral fo $x^-$ in the range of interest, corresponding to ingoing time $y^-$ of order $k^{-1}$:
\begin{align*}
	\partial_- \int_0^{x^-} dt (x^--t)^2 \{u,t\} &= 2\int_0^{x^-}dt (x^--t)\{u,t\} \\
	&\sim \int^{x^-} \frac{x^--t}{(t_\infty -t)^2} dt \\
	&\sim -\log\left(\frac{t_\infty-x^-}{t_\infty}\right)+O(1) \\
	\implies \int_0^{x^-} dt (x^--t)^2 \{u,t\}&\sim \tfrac{2}{k}((\pi T_1 t_\infty)^2-1) + (t_\infty-x^-)\log\left(\frac{t_\infty-x^-}{t_\infty}\right)
\end{align*}
Here we use a late time approximation for $\{u,t\}$, for which the peak of the integrand at $t_\infty-t= 2(t_\infty-x^-)$ gives the logarithm.

We can now assemble the pieces to give the desired variation of the dilaton:
\begin{align}
	 \partial_+ \phi &\sim  -\bar{\phi}_r \left[(2\pi T_1)^2\frac{t_\infty}{t_\infty-x^-} - \frac{k}{t_\infty-x^-} \log\left(\frac{t_\infty}{t_\infty-x^-}\right) \right] \\
	 &\sim -4\pi T_1\bar{\phi}_r \frac{ e^{-\frac{k}{2}y^-}}{t_\infty-x^-}
\end{align}
In the second line, we have used the approximate form \eqref{eq:LateReparam} for the reparameterization to substitute $x^- = f(y^-)$.

To extremize the generalized entropy, we balance this against the lightcone singularity in the entropy as before:
\begin{equation}
	4\pi T_1 \frac{ e^{-\frac{k}{2}y^-}}{t_\infty-x^-} \sim \frac{2k}{x^+-t}
\end{equation}

Now look at the variation of the dilaton parallel to the horizon:
\begin{equation}
	(x^+-x^-)^2 \partial_- \phi = \bar{\phi}_r \left[2-2(\pi T_1 x^+)^2 + k \int_0^{x^-} dt (x^+-t)^2 \{u,t\} \right]
\end{equation}
To analyse the integral, we expand in powers of $t_\infty-x^+$ and use the same approximation for $\{u,t\}$ as before:
\begin{align*}
	\int_0^{x^-} dt & (x^+-t)^2 \{u,t\}-I_\infty \\
	&= (t_\infty-x^+)^2 \int_0^{x^-}\mkern-4mu dt  \{u,t\} -2(t_\infty-x^+) \int_0^{x^-}\mkern-4mu dt (t_\infty-t) \{u,t\} \\
	&\qquad\qquad -  \int_{x^-}^{t_\infty}\mkern-4mu dt (t_\infty-t)^2 \{u,t\} \\
	&\sim \frac{(t_\infty - x^+)^2}{2(t_\infty -x^-)} + (t_\infty -x^+)\log\left(\frac{t_\infty -x^-}{t_\infty}\right) -\tfrac{1}{2}(t_\infty -x^-)
\end{align*}
This matches a result from early times. We can neglect the term quadratic in $t_\infty -x^+$, but the others can be relevant, as seen in the variation of the dilaton:
\begin{align*}
	\partial_- \phi &\sim \frac{\bar{\phi}_r}{(t_\infty- x^-)^2} \left[ 4\pi T_1 (t_\infty-x^+) + k(t_\infty -x^+)\log\left(\frac{t_\infty -x^-}{t_\infty}\right) -\tfrac{k}{2}(t_\infty -x^-)\right] \\
	&\sim \frac{\bar{\phi}_r}{(t_\infty- x^-)^2} \left[4\pi T_1 (t_\infty -x^+)e^{-ky^-/2} -\tfrac{k}{2}(t_\infty -x^-)\right]
\end{align*}

Finally, we compute the variation of the entropy. The $u-y^-$ term is unimportant, but we must include the $\sqrt{f'(y^-)}$ term:
\begin{equation}
	\partial_{x^-} \log f'(y^-) \sim -\frac{1}{t_\infty -x^-}
\end{equation}
From this, we get the variation of the bulk entropy,
\begin{equation}
	\partial_- S \sim \frac{c}{12}\frac{1}{t_\infty - x^-},
\end{equation}
to find
\begin{equation}
	\frac{4G_N}{\bar{\phi}_r}\partial_- S_\text{gen} = \frac{1}{(t_\infty- x^-)^2} \left[4\pi T_1 (t_\infty -x^+)e^{-ky^-/2} +\tfrac{k}{2}(t_\infty -x^-)\right],
\end{equation}
and hence
\begin{equation}
	x^+ - t_\infty \sim  \frac{k}{8\pi T_1}(t_\infty-x^-)e^{ky^-/2}.
\end{equation}
The important term in the variation of the bulk entropy appearing here has a simple interpretation that becomes clearer when rewritten in terms of the variation with respect to $y^-$:
\begin{equation*}
	\frac{\partial S}{\partial y^-} \sim -\frac{c}{12}2\pi T_1 e^{-k y^-/2}
\end{equation*}
Over periods of time small compared to $k^{-1}$, for which the black hole can be regarded as almost static, this variation is approximately constant, and in fact equal to minus the (left-moving half of the) entropy density of the matter CFT at the temperature $2\pi T_1 e^{-k y^-/2}$ of the black hole at the relevant time. This entropy can be thought of as arising from the absence of the ingoing thermal matter that would be present in the Hartle-Hawking state (for which the entropy is constant by $SL(2)$ invariance), but has instead escaped into the bath.

Putting this together with extremization in the $x^+$ direction, we can locate the quantum extremal surface:
\begin{align}
	x^+-t_\infty &\sim \frac{t_\infty-t}{3} \\
	t_\infty -t &\sim \frac{3k}{8} t_\infty  e^{\frac{k}{2} y^-} (t_\infty -x^-)
\end{align}

At this ingoing time $x^-$, we have
\begin{equation}
	t_\infty -x^+ \sim \begin{cases}
		-\tfrac{1}{3}(t_\infty-t) & \text{(QES)} \\
		0 & \text{(event horizon)} \\
		\tfrac{1}{3}(t_\infty-t) & \text{(apparent horizon $\partial_-\phi=0$)}
	\end{cases}
\end{equation}
so the QES is the same distance inside the event horizon as the apparent horizon is outside.

Expressing this  entirely in terms of the boundary proper time $u$ and corresponding ingoing coordinate $y^-$, we have our final location for the quantum extremal surface:
\begin{equation}\label{eq:lateQES}
	u \sim y^- + \frac{e^{\frac{k}{2}y^-}}{2\pi T_1 }\log\left(\frac{8 \pi T_1 e^{-\frac{k}{2}y^-}}{3k}\right)
\end{equation}
In the case that this is the quantum extremal surface of minimal generalized entropy (of which more in a moment), this gives a beautiful quantum geometric realization of the Hayden-Preskill experiment. The quantum extremal surface demarks the region of spacetime to which the system has access. This region is bounded to the future by the event horizon\footnote{This is the outside of the horizon in the event that evaporation continues to very late times.  However, if we stop evaporation at some earlier time, the event horizon will be further out, due both to the necessary positive energy shock from turning off the coupling and the Hawking radiation that no longer is allowed to escape. Indeed, from \eqref{eq:lateQES} we see that the QES lags a scrambling time behind the boundary time so that an ingoing $O(1)$ positive energy pulse from switching off the coupling has a large effect on its location relative to the final future event horizon.}, and to the past by the ingoing time $y^-$ solving \eqref{eq:lateQES}. If we throw in a message which falls behind the horizon before this time its information is lost to the system, and obversely, is retrievable from the Hawking radiation in the bath, combined with the purifying system (the other side of the black hole). The time delay apparent in \eqref{eq:lateQES} is precisely the scrambling time, with $\beta = \frac{e^{\frac{k}{2}y^-}}{T_1 }$ being the thermal scale at that time. Indeed, this delay may be written
\begin{equation}
\label{eq:tHPdelay}
t_{HP} = \frac{\beta}{2\pi} \log \left[\frac{16}{c}(S-S_0) \right],
\end{equation}
with $S$ the density of states at time $u$ on the boundary.  This result is precisely of the form \eqref{thp1} with coefficient $\alpha_{HP}=1$ up to the interesting-but-tiny correction $\frac{\beta}{2\pi} \log \left(\frac{8}{c}\right)$.
We find that the scrambling time is slightly reduced large $c$, because more bulk degrees of freedom encode proportionally more information in the Hawking radiation.  One is tempted to associate this correction with a $c$-dependent minimal $\delta E$ in \eqref{tscr1}, but this remains to be understood in detail.

We now compute the generalized entropy associated to these surfaces. Since the QES is very close to the event horizon, to leading order it suffices to evaluate the dilaton on the horizon $x^+=t_\infty$ by the integral \eqref{eq:dilatonInt}:
\begin{equation}
	I(x^+=t_\infty, x^-) = \int_0^{x^-} dt (x^--t)(t_\infty -t) \{u,t\}
\end{equation}
To do this, we use the same approach as before, using the fact that it approaches $I_\infty$ as $x^-\to t_\infty$, and estimate the derivative at late times:
\begin{align*}
	\partial_- I(t_\infty,x^-) &= \int_0^{x^-} dt (t_\infty-t)\{u,t\} \\
	& \sim \int^{x^-} \frac{dt}{2(t_\infty -t)}\\
	&\sim -\tfrac{1}{2} \log(t_\infty -x^-) \\
	\implies  I(t_\infty,x^-) &\sim I_\infty +\tfrac{1}{2}(t_\infty-x^-) \log\left(\frac{t_\infty -x^-}{t_\infty}\right)
\end{align*}
This gives us the dilaton value on the horizon, where the quantum extremal surface is located:
\begin{align}
	\phi &\sim \bar{\phi}_r \left[2\pi T_1 + \tfrac{k}{2}\log\left(\frac{t_\infty -x^-}{t_\infty}\right)\right] \\
	&\sim \bar{\phi}_r  2\pi T_1 e^{-\frac{k}{2}y^-}
\end{align}
This gives the unsurprising result that the dilaton at the horizon matches the thermodynamic entropy associated with the energy at the corresponding ingoing time. As a further check, it is also consistent with the earlier result for the $\partial_-$ variation of the dilaton. The bulk entropy gives a subleading contribution to $S_\text{gen}$; its gradient competes with the area term, but the quantity itself does not.

We note that an alternative way to perform the analysis at late times is to use the AdS isometries to change to a set of coordinates adapted to the time in question, which we discuss briefly in section \ref{sec:disc}.

\begin{figure}
\begin{center}
	\includegraphics[height=7cm]{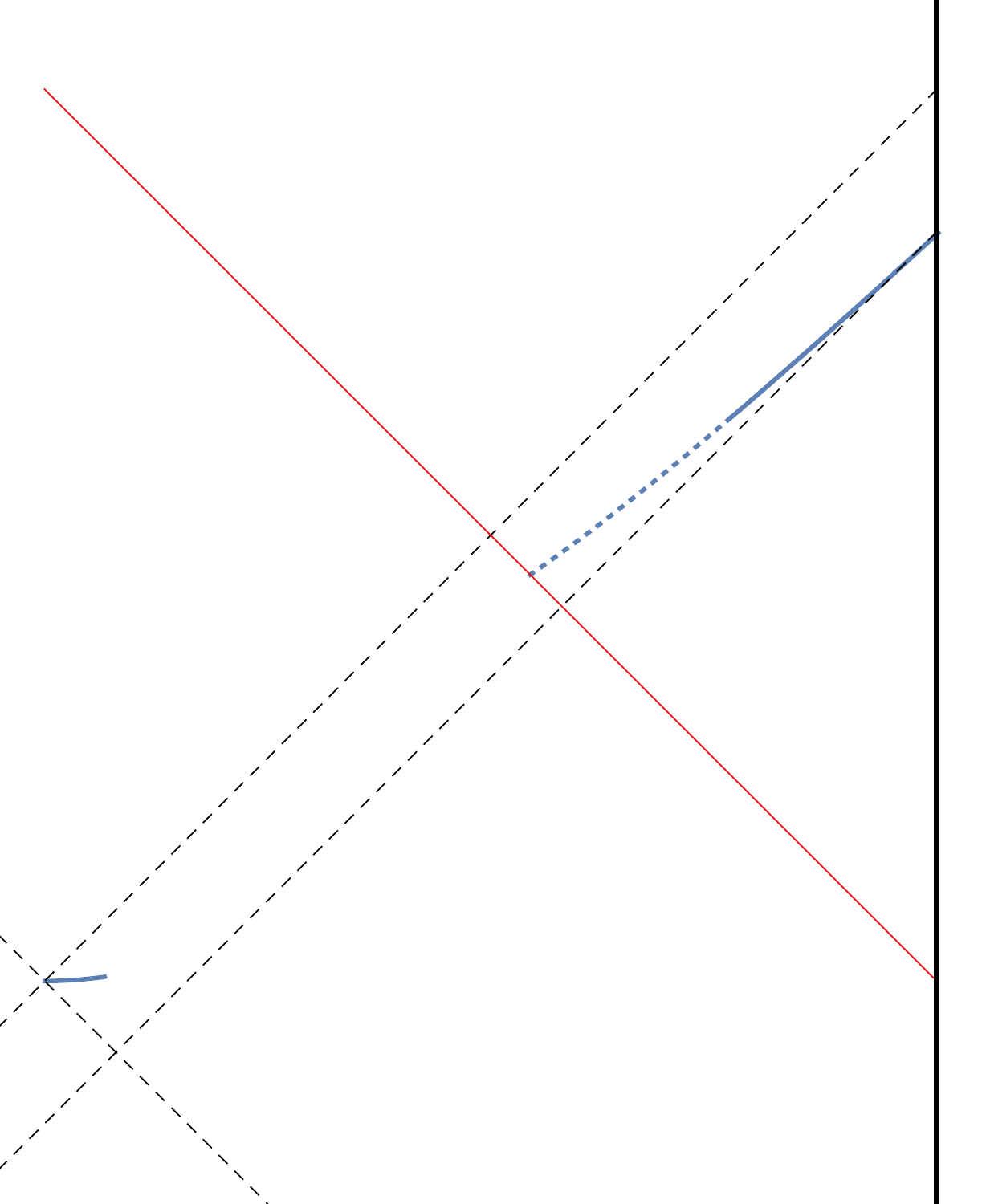}
	\end{center}
	\caption{The location of quantum extremal surfaces as boundary time evolves (solid blue curves). The dashed lines indicate the original horizon before coupling to the bath, and the final event horizon. The red line indicates the shock at $x^-=0$. The dotted blue line is the non-minimal quantum extremal surface before the Page time phase transition.}
\label{fig:QESDiagram}
\end{figure}

The quantum extremal surface of section \ref{sec:earlyQES}, which dominates at early times, gives a linearly increasing generalized entropy starting close to the entropy of the original black hole, with temperature $T_0$. The new surface we have found has genralized entropy which decreases linearly, but starting close to the thermodynamic entropy associated to the perturbed black hole after the shock, of temperature $T_1$. The latter is initially larger, but at a time of order $t_P\sim k^{-1}\frac{T_1-T_0}{T_0}$ they exchange dominance, and our new surface becomes the quantum extremal surface. We interpret this phase transition with the Page time\footnote{A.A. thanks G. Penington for discussion of this point.}, when the black hole once again becomes old; see figure \ref{fig:QESDiagram}.

With these two families of quantum extremal surfaces, the generalized entropy perfectly reproduces the Page curve, indicative of unitary evaporation. The entropy increases at early times, since the outgoing Hawking radiation is entangled with the remainder of the system. This continues until the entropy of the black hole is close to its maximum at the given energy, at which point the entropy can only reduce, implying that the radiation is purified by early radiation (here including the left side black hole).

It is striking, and perhaps surprising, that we have recovered the Page curve from a semiclassical calculation. However, we emphasize that this does not resolve the information paradox, since it supposes that the generalized entropy of the quantum extremal surface correctly captures the fine-grained entropy of the black hole, without explaining the required correlations between early and late Hawking radiation. Indeed, the state of the matter in the bath at late times does not contain such correlations, and calculating the entropy of the bath from \eqref{eq:bathS} points towards this loss of unitarity:
\begin{equation}
\label{eq:Sbath}
	S_\text{bath} \sim \frac{\bar{\phi}}{4G_N}4\pi T_1 (1-e^{-\frac{k}{2}u})
\end{equation}
This increases for all time, at the rate given by the current temperature, approaching the entropy of a black hole at temperature $2T_1$.

Due to the large ground state entropy of $R$, there is no immediate contradiction between \eqref{eq:bathS} and unitarity. In particular, the Araki-Lieb inequality is always satisfied. However, $S_\mathrm{bath}$ can be made arbitrarily large by successive iterations of injecting additional pulses of low-entropy energy into the black hole and further evaporation into the bath. In this way one demonstrate information loss as a violation of the Araki-Lieb inequality while remaining in the near-extremal regime throughout the evaporation. We will discuss such issues further in section \ref{sec:disc}.

\subsection{Summary}
\label{sec:sum}

Since the final discussion in section \ref{sec:disc} will focus on more conceptual issues, we now pause to summarize the above technical results.  In doing so, it is useful to take the viewpoint that begins with 3 systems $L, R, B$ which at first do not interact.  Here $L,R$ are both AdS$_2$ black holes, and the black holes are highly entangled.  The bath $B$ begins in its ground state.  This initial state is then perturbed by turning on a coupling between $R$ and $B$, localized at the boundary of both systems.  The coupling is then left on and becomes time-independent after a short initial transient.

The primary object of our study was the location of quantum extremal surfaces defined by the bulk entropy between an interior point and a point on the cut-off boundary at physical time $u$.  Adding this entropy to $\frac{\phi_0 + \phi}{4G}$ gives the generalized entropy, and for each $u$ any QES are identified by extremizing the result with respect to the internal point.  Our results then concern the evolution of such QES with $u$.

Before the coupling is turned on, there is a unique QES that (due to symmetry) agrees precisely with the classical extremal surface at the bifurcation surface of the original black hole horizon.  After the coupling is turned on, this QES moves out toward the boundary in a spacelike direction.  Several transient effects then occur whose effects on the entropy of the system were shown in figure \ref{fig:EarlySgen}.  These effects are easily understood in terms of bulk quantum field theory with perturbative gravitational back-reaction.  Switching on the coupling causes an initial sharp increase in entropy due to the production of short-distance entanglement between $R$ and $L$.  This is then followed by a decrease in entropy due to the escape of thermal radiation from $R$ into the bath $B$.  However, after a scrambling time $\frac{\beta_1}{2\pi}\log \left(\frac{E(T_1)}{E_S} \right)$ \eqref{eq:tSmin} one finds that the effective horizon of the black hole (say, the apparent horizon) has moved even further outward.  As a result, the entropy defined by this QES begins to increase again as Hawking particles are emitted into the bath and their partners traveling inward add entropy to $R$.  The resulting linear increase in entropy corresponds to the rising early entropy in the familiar Page curve of \cite{Page:1993df}.

The entropy associated with this QES in fact continues to rise for all time. The corresponding entropy thus tracks precisely what one would expect if information were lost in evaporating black holes.  However, since the actual entropy is described by the QES with the smallest $S_{gen}$, this same increase means that the dominant QES is likely to experience a phase transition at late times if we can identify another QES with smaller time derivative of $S_{gen}$.

Such a new QES can indeed be found emerging from the positive-energy shock that was produced at the AdS$_2$ boundary by turning on the coupling between $R$ and $B$ and which then falls into the black hole.  The shock heats the black hole to a higher temperature $T_1$, associated with a higher energy $E_1$ and a higher Bekenstein-Hawking entropy $S_1$.  The new QES is not a small perturbation to a classical extremal surface, but instead arises because there are surfaces in the classical geometry with sufficiently small expansions that they can be turned into a QES by quantum effects.  Since it sits close to the new apparent horizon associated with the increased energy of the black hole, its entropy begins close to $S_1$, which at first significantly exceeds that of the first QES described above.  This means that at the early times shown in figure \eqref{eq:tSmin}, the new QES has larger entropy and thus is not relevant to determining the entanglement wedge.

However, as the back hole continues to radiate and lose energy to $B$, the new QES remains close to the apparent horizon and thus slowly decreases in entropy.  In particular, the actual value of $S_{gen}$ at the new QES remains very close to that of the classical term $\frac{\phi_0 + \phi}{4G}$, and thus close to the Bekenstein-Hawking density of states associated with the (time-dependent and now decreasing) energy of $R$.  In other words, the $S_{gen}$ of this QES agrees well with the entropy one would expect if one began with a maximally-entangled black hole and then watched it slowly evaporate via a unitary process.  It is thus clear that at the relevant notion of the Page time the $S_{gen}$ of this new surface will become less that of the other QES described above (whose $S_{gen}$ continues to increase).  This gives the stated phase transition of the dominant QES and then reproduces the decreasing part of the Page curve of \cite{Page:1993df}.

It is also useful to note that the 2nd QES moves outward in a spacelike direction as time passes, though its motion becomes asymptotically null as the black hole reaches its final extremal equilibrium.  The motion of both QES was shown above in figure \ref{fig:QESDiagram}.  The worldline of the novel QES is marked as dotted in the region where it fails to dominate, and then solid where it dominates after the phase transition.  This figure also shows the novel QES approaching the final event horizon from the interior at late times.

What is not shown in figure \ref{fig:QESDiagram} is the relation between the physical boundary time $u$ and the location of the corresponding QES on the solid blue curve.  However, as discussed around equation \eqref{eq:tHPdelay} this relation becomes quite simple at late times.  In that limit, the dominant QES for boundary time $u$ lies on the ingoing null geodesic that was emitted from the boundary at a time $u-t_{HP}$ with
\begin{equation}
\label{eq:tHPdelay2}
t_{HP} = \frac{\beta}{2\pi} \log \left[\frac{16}{c}(S-S_0) \right].
\end{equation}
This is a holographic manifestation of the Hayden-Preskill effect, where an ingoing signal disappears from the entanglement wedge of $R$ after the time \eqref{eq:tHPdelay2} and enters the entanglement wedge that would be defined by merging $L$ and $R$; i.e., the signal can be recovered from the part of the system that describes the complement of the black hole into which the signal has fallen.

\section{Left and Right Quantum Extremal Surfaces}
\label{sec:gap}

For bulk matter in a pure state, a QES of a boundary region is identical to the QES of its complement. In particular, in a two-boundary geometry, the QES of the complete left boundary is identical to the QES of the complete right boundary whenever the full bulk state is pure.

In our situation, where the two-sided black hole in question has emitted some radiation into the bath, however, the bulk state is mixed: the right and left QESs are no longer required to coincide. Put differently, complementary recovery is not guaranteed when the bulk state is not pure~\cite{Harlow:2016vwg}.  This discrepancy between left and right QESs, which in turn corresponds to the gap between the left and right entanglement wedges, is an important aspect of our holographic realization of the Hayden-Preskill protocol.

\begin{figure}
\center
\includegraphics[width=0.5\textwidth]{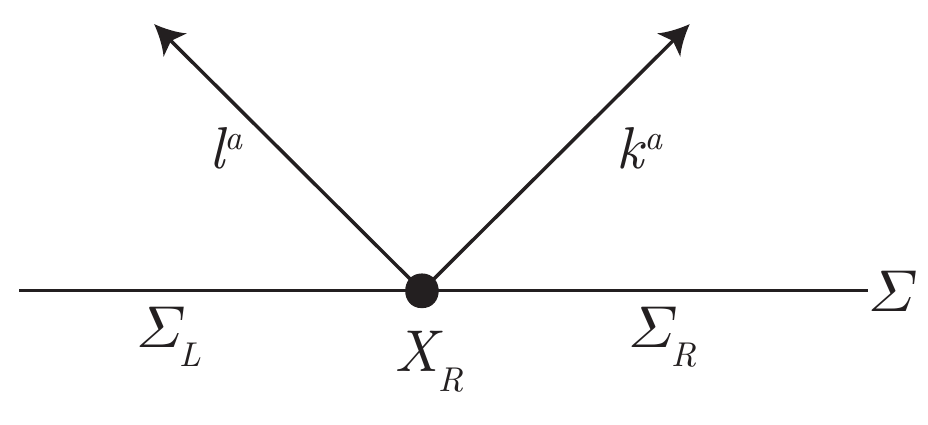}
\caption{A Cauchy slice $\Sigma$ containing the right quantum extremal surface $X_{R}$. By assumption, $X_{R}$ splits $\Sigma$ into two sides, labeled $\Sigma_{R}$ and $\Sigma_{L}$. $\Sigma_{R}$ is a Cauchy slice of the entanglement wedge of the right side, while $\Sigma_{R}$ is not necessarily a Cauchy slice of the left entanglement wedge. The null vectors $\ell^{a}$ and $k^{a}$ are normal to $X_{R}$. While this cartoon is two-dimensional, this setup is not restricted to any particular number of dimensions. }
\label{fig:decomp}
\end{figure}

Let us therefore investigate the left quantum expansion of the right QES in a general setting. We consider a generic bulk state in a two-sided (or in principle multi-boundary) geometry. We will temporarily work in general bulk dimension $d\geq 2$ and with bulk matter described by any local (not necessarily conformal) quantum field theory. Assuming, as we have thus far in this paper, that the right (minimal entropy) QES $X_{R}$ lies in a globally hyperbolic region~\footnote{Here we mean the appropriate generalization to global hyperbolicity for asymptotically AdS spacetimes~\cite{Ger70}.} which is well-described by an approximate geometry, we consider a Cauchy slice $\Sigma$ of the bulk spacetime containing $X_{R}$. The right QES $X_{R}$ splits $\Sigma$ into two components: $\Sigma_{L}$, the left of $X_{R}$ on $\Sigma$, and $\Sigma_{R}$, the right of $X_{R}$ on $\Sigma$; see Fig.~\ref{fig:decomp}. This allows us to define the right and left reduced density matrices for bulk quantum fields: $\rho_{\Sigma_{R}} = \mathrm{tr}_{\Sigma_{L}}\rho$ and $\rho_{\Sigma_{L}} = \mathrm{tr}_{\Sigma_{R}}\rho$, which in turn define the right- and left- bulk von Neumann entropies of $X_{R}$: $S_{R}[X_{R}]$ and $S_{L}[X_{R}]$.

Because $X_{R}$ is a right QES, it is by definition a stationary point of the generalized entropy functional: the variation of $S_{R}[X_{R}]$ due to shape deformations of $X_{R}$ vanishes to first order in the deformation. Equivalently, the right quantum expansion of $X_{R}$ in any null direction vanishes. If $\ell^{a}$ and $k^{a}$ are linearly independent null normals to $X_{R}$ (in our $d=2$ setup above, these may be chosen to be e.g. along $x_{\pm}$) with affine parameters $\lambda_{\ell}$ and $\lambda_{k}$ respectively, we may schematically write this in the following way:
\begin{subequations}
\begin{align}
& \frac{\delta S_{\mathrm{gen}}^{(R)}[X_{R}]}{\delta \ell^{a}}=0\\
& \frac{\delta S_{\mathrm{gen}}^{(R)}[X_{R}]}{\delta k^{a}}=0,
\end{align}
\end{subequations}
where $S_{\mathrm{gen}}^{(R)}[X_{R}]$ is the right generalized entropy of $X_{R}$. Thus the right generalized entropy of $X_{R}$ does not change to first order when $X_{R}$ is infinitesimally deformed in a null direction (or, as noted above, any direction; it will simply turn out to be useful to work with null vectors).

To compare the left- and right- quantum expansions of $X_{R}$ (and thus ascertain whether $X_{R}$ is a left QES, and if not, by how much it fails to be one), consider deforming $X_{R}$ along the $k^{a}$ or $\ell^{a}$ directions. Since the change in area or in the dilaton for $d=2$ is independent of whether we evaluate it on the left or right side, the difference between the left and right quantum expansions is entirely due to the evolution of $S_{R}[X_{R}]$ and $S_{L}[X_{R}]$ along the null directions.

\begin{figure}
\center
\includegraphics[width=0.5\textwidth]{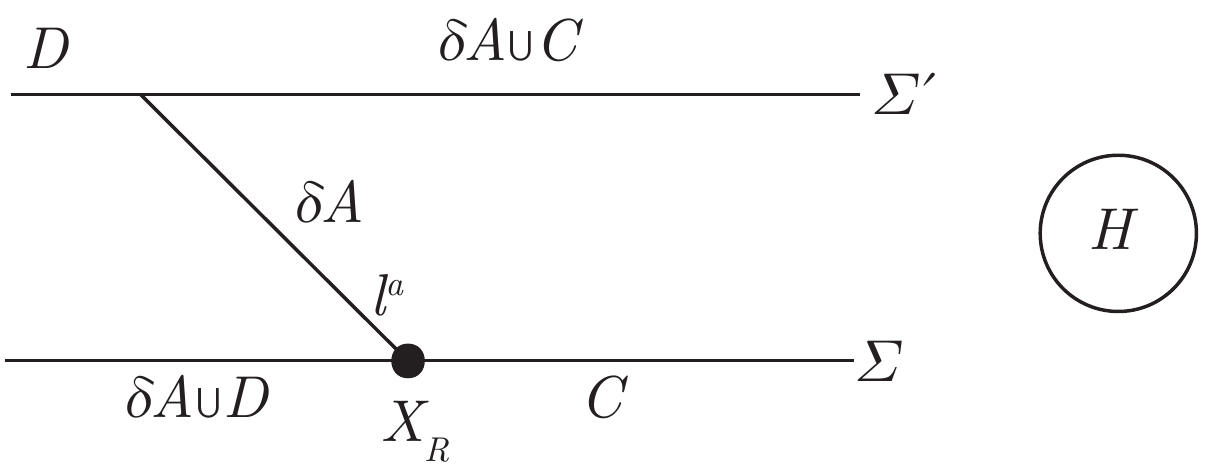}
\caption{Here $\Sigma_{R}=C$ and $\Sigma_{L}=\delta A \cup D$. The infinitesimal deformation along $\ell^{a}$ is labeled $\delta A$ and is parametrized by a choice of affine parameter $\lambda_{\ell}$ along the $\ell^{a}$ congruence. The $\ell^{a}$ congruence defines a similar split of the future Cauchy slice $\Sigma'$ into the right side, $\delta A\cup C$ and the left side $D$. }
\label{fig:evolution}
\end{figure}

To track this evolution, we evolve to a new Cauchy slice $\Sigma'$ to the future of $\Sigma$, and divide $\Sigma'$ into left and right components using $\ell^{a}$ or $k^{a}$ This is illustrated in Fig.~\ref{fig:evolution} for  $\ell^{a}$.

Using the decomposition labeled in Fig.~\ref{fig:evolution} in the $\ell^{a}$ direction as well as the fact that $X_{R}$ is a  \textit{right} quantum extremal, we know that as we take $\delta A$ to zero, parametrized by the affine parameter $\lambda_{\ell}$ along $\ell^{a}$, we get:
\begin{equation}
\delta S_{R}[X_{R}]=S[C]-S[C\cup \delta A] =\mathcal{O}(\lambda^{2}).
\end{equation}
To relate the change along $\lambda_{\ell}$ of $S_{L}$ with that of $S_{R}$, we would like to compare $S[D\cup \delta A]-S[D]$ with $S[C]-S[C\cup \delta A]$. This can be done with a judicious application of strong subadditivity (SSA)~\cite{LieRus73} to the system consisting of $\{\delta A, C, H\}$, where $H$ is the emitted radiation:
\begin{equation} \label{eq:SSA}
S[C \cup H]+S[C\cup \delta A] -S[\delta A\cup C\cup H]-S[C] \geq 0.
\end{equation}
Using the fact that $\rho_{\delta A \cup C \cup H \cup D}$ is pure, this immediately yields the desired comparison:
\begin{equation}
S[D\cup \delta A]-S[D]\geq S[C]-S[C\cup\delta A]=\mathcal{O}(\lambda^{2}).
\end{equation}
So we find that
\begin{equation} \frac{\delta S_{\mathrm{gen}}^{(L)}[X_{R}]}{\delta \ell^{a}}\geq \frac{\delta S_{\mathrm{gen}}^{(R)}[X_{R}]}{\delta \ell^{a}}=0,\end{equation}
with the opposite inequality for deformations along the $k^{a}$ direction:
\begin{equation} \frac{\delta S_{\mathrm{gen}}^{(L)}[X_{R}]}{\delta k^{a}}\leq \frac{\delta S_{\mathrm{gen}}^{(R)}[X_{R}]}{\delta k^{a}}=0.\end{equation}
That is, $X_{R}$ is, from the perspective of the left side, a quantum ``untrapped'' (also called ``normal'') surface unless the SSA~\eqref{eq:SSA} is saturated.

When is SSA saturated? It is saturated strictly when $\rho_{\delta A\cup C\cup H}$ is a so-called Markov state: then the full state can be recovered completely from either of its marginals $\rho_{\delta A \cup C}$ or $\rho_{C\cup H}$. In particular, there exists a (state-dependent) \textit{recovery map} $\mathbb{I}_{A}\otimes{\cal R}_{C}$, where $\mathbb{I}_{A}$ is the identity on $A$ and ${\cal R}_{C}$ is a map from states on $C$ to states on $C\cup H$, such that~\cite{HayJoz04}:
\begin{equation}
\mathbb{I}_{A}\otimes{\cal R}_{C}[\rho_{\delta A\cup C}]=\rho_{\delta A \cup C\cup H}.
\end{equation}
That is, the right QES is a left QES whenever the radiation is sufficiently entangled with the right entanglement wedge that we can fully reconstruct (in the sense of the recovery map above) the state of the radiation by just knowing the state of the entanglement wedge. This is precisely the case e.g.  when we throw the radiation back into the left CFT so that the full bulk state is pure: the two QESs coincide; there is no gap between the left and right entanglement wedges.

It thus seems reasonable to anticipate that when the right QES is very close to also being a left QES -- such as might be the case when the two surfaces in question are close to one another~\footnote{Note that approximate extremality does not guarantee proximity of extremal surfaces: the left quantum expansions of $X_{R}$ being close to zero is not enough to guarantee that there is a neaby left quantum extremal surface.} -- that $\rho_{\delta A \cup C\cup H}$ would be very close to a Markov state by some measure. That is, we may expect that the radiation is sufficiently entangled with the right entanglement wedge that we may approximate its state using the state-dependent maps above.

In the context of type I von Neumann algebras, this can be made precise. The failure of saturation of strong subadditivity is bounded from below~\cite{FawRen14, SutFaw15}:
\begin{equation}\label{eq:FRBound}
S[C\cup H]+S[C\cup \delta A] -S[\delta A\cup C\cup H]-S[C] \geq -2 \ln \left (\sup \limits_{{\cal R}_{C}}F \left [\rho_{\delta A \cup C \cup H} | \mathbb{I}_{A}\otimes{\cal R}_{C}(\rho_{\delta A\cup C}) \right]\right).
\end{equation}
Here $F$ is the fidelity between the state $\rho_{\delta A\cup C\cup H}$ and a state we obtain from a recovery map acting on $\rho_{\delta A\cup C}$, $\mathbb{I}_{A}\otimes{\cal R}_{C}(\rho_{\delta A\cup C})$; the supremum is taken over all recovery maps (this is sometimes termed ``fidelity of recovery''~\cite{BerTom15, SesWil15}). This suggests that if our right QES is close to being a left QES, then our state is very well approximated by a Markov state, the radiation $H$ is entangled mostly with the right entanglement wedge, and there exists a state-dependent recovery map giving a good approximation of $H$ from $C$. More precisely, $\rho_{\delta A\cup H}$ is close (as measured by the fidelity) to a product state.

\begin{figure}
\center
\includegraphics[width=0.5\textwidth]{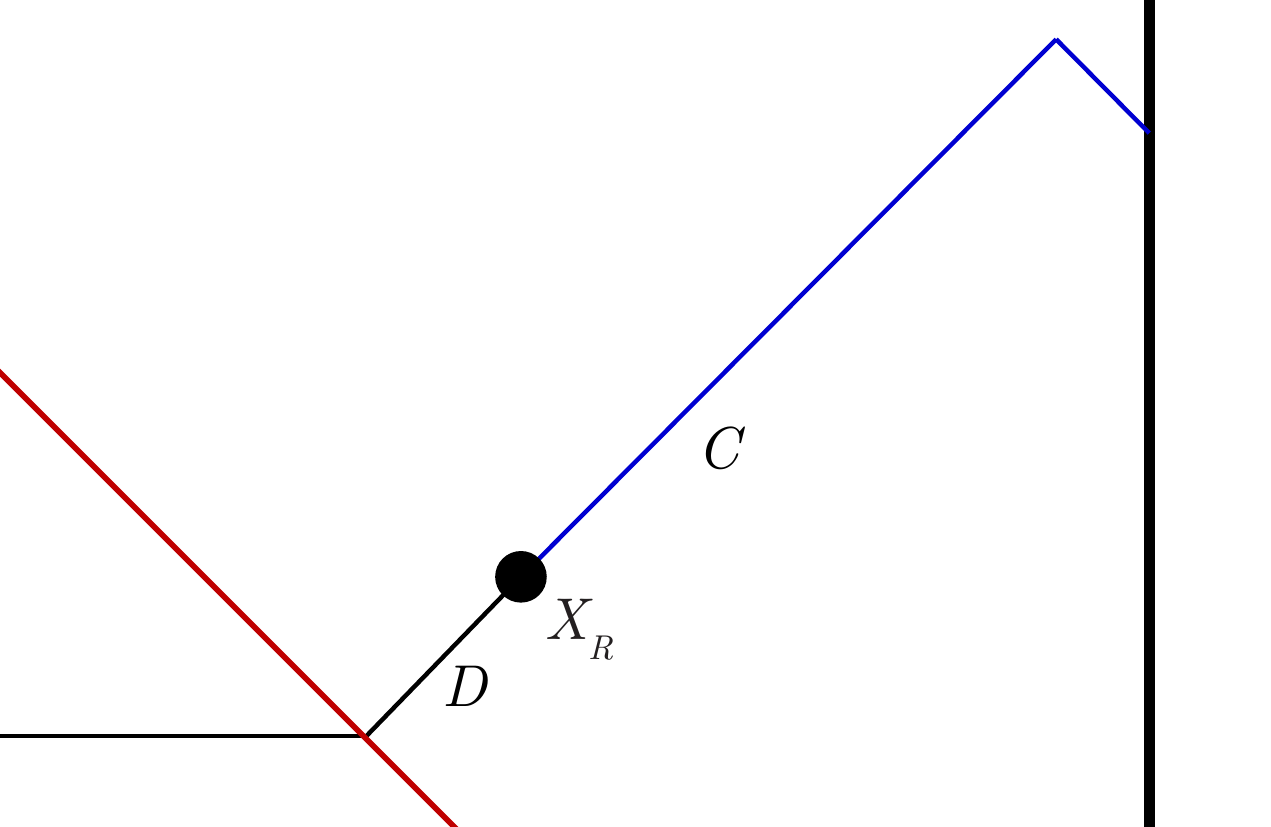}
\caption{Breaking up the right interval and left intervals into null components for the calculation of the change in the entropy of $X_{R}$ with variations along $y_{-}$. The red line is the shell.}
\label{fig:LeftVsRight}
\end{figure}

Thus far our discussion in this section has been for general holographic systems. Let us now carry out the explicit calculation comparing the left and right quantum expansions of the right QES for our $D=2$ black hole with conformal matter (we will see that SSA is very far from saturation in our system). For simplicity, we focus on the late (but not very late) time region where the (minimal entropy) right QES $X_{R}$ lies entirely to the future of the shock (as illustrated in Fig.~\ref{fig:LeftVsRight}). To evaluate the left quantum expansions in the $x^{+}$ and $y^{-}$ directions, consider a partial Cauchy slice extending from $X_{R}$ to the left boundary. Unitarity of the bulk theory allows us to wiggle the Cauchy slice in any way we want, and for convenience we break it up into a null, constant $x_{+}$ component to the future of the shell and a component to the past of the shell. By Eq.~\ref{eq:constantentropy}, the renormalized entropy of intervals to the past of the shell in the Poincar\'e vacuum is independent of the interval, so it does not contribute to the derivative of $S_{L}[X_{R}]$ in either the $x^{+}$ or $y^{-}$ directions.

We first consider the left- and right- quantum expansions in the $y^{-}$ direction, which in the discussion above corresponds to the $k^{a}$ direction. This calculation is particularly simple, since the the right interval can be broken into two null intervals: from $x_{b}=f(y_{b})$ to $(x^{+}, y^{b})$ and from $(x^{+}, y^{b})$ to $(x^{+}, y^{-})$, the coordinates of $X_{R}$. The interval of interest in for the left quantum expansion, which we shall call $D$, extends from $(x^{+},0)$ to $(x^{+},y)$. This is illustrated in Fig.~\ref{fig:LeftVsRight}. Using ~\eqref{eq:entropy2}, we immediately obtain the following expressions for the difference in the expansion along $y^{-}$ is:
\begin{equation}
\partial_{y^{-}} S_{\mathrm{gen}}^{(R)}[X_{R}] -\partial_{y^{-}} S_{\mathrm{gen}}^{(L)}[X_{R}]= \frac{cy_{b}}{12y (y_{-}-y_{b})}< 0,
\end{equation}
where the inequality follows from the fact the by assumption the right QES lies to the future of the shock, and $0<y^{-}<y_{b}$ (as a sanity check, this is consistent with the sign expected from strong subadditivity). We note that as desired, the inequality is \textit{not} saturated: the right QES fails to be a left QES (in fact, the difference is large). As explained above, we may interpret this as a manifestation of the fact that the radiation is far less entangled with the right half of the black hole than what would be necessary for us to recover even approximately the state of the radiation purely from knowing the state on this side of the black hole.

For completeness, we will also compute the difference in left- and right- quantum expansions in the $x_{+}$ direction, which in the above conventions corresponds to the $\ell^{a}$ direction. Using the same decomposition of the left interval as above, we obtain:
\begin{equation}
\partial_{x_{+}} S_{\mathrm{gen}}^{(R)}[X_{R}] -\partial_{x_{+}} S_{\mathrm{gen}}^{(L)}[X_{R}]= \frac{c}{12(x-x_{b})}>0.
\end{equation}
Again the inequality (which has the correct sign dictated by strong subadditivity) fails to saturate by a large amount, as expected by the considerations above.

Finally, we have here used the fact that the failure of the right QES to achieve left quantum extremality is parametrized by the failure of a state to be Markovian; this is bounded from below by the failure of recovery maps on the right entanglement wedge to approximate the state of the radiation (where this failure is measured by the fidelity). It is tempting to use this non-saturation of strong subadditivity for generic states to attempt to prove that \textit{in general} right and left QESs do not coincide. However, this is where we must be careful with the discreteness of the bound in~\eqref{eq:FRBound}. To prove that~\eqref{eq:SSA} is not saturated, we must show that the failure of SSA to saturate scales no faster with $\lambda$ than $\cal{O}(\lambda)$ as we shrink $\delta A$ to zero. If it goes to zero any faster, the first derivative -- i.e. the quantum expansion -- will not be sensitive to it. The bound~\eqref{eq:FRBound} applies to discrete systems and thus contains no immediate information on a continuously shrinking region. It is only indicative of the statement that the inequality is not, generically, saturated to \textit{all} orders in $\lambda$. Even in the discrete model, it is unclear to us whether the nonsaturation will be linear in $\lambda$ (though the fact that the sign of the inequality flips when $\lambda \rightarrow -\lambda$ does suggest that the a linear scaling is consistent). It may be possible to use different techniques to find the scaling order; we leave this to future work.

\section{Discussion}
\label{sec:disc}

The main outcome of our work was to show that Quantum Extremal Surfaces (QES) accurately described the expected unitary evaporation of black holes. While we focused on a particular model in low dimensions, we will shortly describe why a similar story must occur in much more general contexts.  Crucially, we identified a novel QES that cannot be described as a small quantum correction to a classical extremal surface.  Instead, this QES arises only due to quantum effects. We will return to the apparent tension in this statement below, but for the moment comment only that this new QES begins to dominate at the Page time, and that the corresponding quantum-HRT phase transition is directly responsible for the fact that our entropy decreases after the Page time as predicted by unitarity. See also section \ref{sec:sum} for a succinct summary of our technical results.

We now reiterate these results with an eye toward interesting conceptual issues and future directions.  The objective of sections \ref{sec:setting}-\ref{sec:QES} was to track the QES under one-sided evaporation of a two-sided thermofield-double-like black hole with initial temperature $T_0$.  At some finite time we turned on a coupling that allowed the right side $R$ of our system to freely radiate into a bath $B$. Although we worked only at the level of perturbative quantum corrections to classical solutions, the QES behaved in ways consistent with general expections from non-perturbative unitary evolution.  In particular, the initial switching on of the coupling induces correlations between $R$ and $B$ which increase the entropy of $R$ and also $S_{gen}$ of the right QES by an amount $S_{S}$.  The switching operation also injects energy $E_{S}$ into our black hole, with $E_{S} \gg T_0$ so that the density of states of $R$ now significantly exceeds $S_{\text{gen}}$ of the QES.  But this new energy takes some time to scramble, so as shown in figure \ref{fig:EarlySgen}  for a scrambling time (with coefficient $\alpha_{S}=1$ the system proceeds as if this new phase space had not opened up.  In particular, $S_{\text{gen}}$ at the QES decreases for a time $\frac{\beta_0}{2\pi} \ln \frac{E}{E_S}$  \eqref{eq:tSmin} as expected from \eqref{tscr1} with coefficient $\alpha_{S}=1$),  and then increases again as the emission of Hawking radiation after this scrambling time creates more entanglement between $R$ and $B$.

Further emission of Hawking radiation continues to cause
the generalized entropy of the QES to increase while the energy of $R$ and thus its density of states decreases.  Unitarity would require this behavior to change at what we may call the Page time for this setup  when the two become equal.  Thus far the QES itself has moved continuously in a spacelike but nearly-null direction\footnote{The fact that the QES moves in a spacelike or nearly-null direction towards the boundary $R$ provides an important consistency condition for the case at hand where the extracted energy from $R$ is immediately deposited in $L$ as it precludes any bulk causal signaling from the left entanglement wedge to the right entanglement wedge. }. But at this Page time there is a phase transition such that continuous deformations of the original QES no longer have minimal $S_{\mathrm{gen}}$.

Instead, another family of QESs becomes minimal after the Page time.  The new family lies closer to the boundary than the ingoing pulse of energy $E_S$; see figure \ref{fig:QESDiagram}.  In this new family of QESs, $S_{\text{gen}}$ is a {\it decreasing} function of time that largely tracks the density of states of $R$ as determined by its energy.  Indeed, in the asymptotic future, $S_{\text{gen}}$ becomes equal to the ground-state entropy of $R$ up to a term associated with the coupling to $B$.  Furthermore, this QES lies close to the future event horizon of $R$, and has an ingoing null coordinate that at late time lags that of the time-evolving boundary by precisely $t_{HP}= \frac{\beta}{2\pi} \ln \left[ \frac{16}{c}(S-S_0)\right]$ \eqref{eq:tHPdelay}, where $S$ is the density of states of $R$ at the given time and $S_0$ is the ground state entropy.  As described in section \ref{sec:HoloHP}, such a lag is to be expected from the holographic description in \cite{Almheiri:2018xdw} of the Hayden-Preskill protocol \cite{Hayden:2007cs}.  Note that the above scrambling time is not just as a rough timescale in our calculation, but as a sharp threshold. In particular, we find the coefficient to be $\alpha_{HP}=1$ in \eqref{thp1}. We may thus say that it gives the Hayden-Preskill delay time for our system in the limit of small messages.

It is tempting to associate the correction $\frac{\beta_0}{2\pi} \ln \frac{16}{c}$ to \eqref{thp1} with a $c$-dependent minimum $\delta E$ in \eqref{tscr1}, but this remains to be understood in detail.  It would also be very interesting to reproduce both this term and the coefficients $\alpha_{HP}=1$ and $\alpha_{S}=1$ from a calculation in an SYK model \cite{Kitaevtalk, Maldacena:2016hyu, Sachdev:1992fk} or the like.  It would also be interesting to compare our results with analogous calculations for higher dimensional black holes.  On general grounds one expects similar behavior, but one would like to understand the extent to which the coefficients $\alpha_{S \min} = \alpha_{HP}=1$ and $\frac{16}{c}$  found here are universal and the extent to which they depend on properties of the black hole.

The fact that bulk entropy, which for us generally does not exceed $O(\ln G_N)$, could significantly affect the location of the QES was due to the presence of large gradients in this entropy.  Such large gradients arise naturally in the context of evaporating black holes from the well-known large boosts.  It would be interesting to understand if such large gradients in entropy might somehow invalidate the semiclassical approximation used here, though we see no immediate reason for this to be the case\footnote{We thank D. Harlow for discussion on this point.}. Indeed, the size of such gradients is coordinate dependent, and with an appropriate choice of outgoing coordinate, the gradient of the entropy is of order one, and the gradient of the dilaton is suppressed by a factor of $G_N$.

This occurs naturally if we choose coordinates adapted to the boundary time of interest, in which case gradients of the dilaton a scrambling time in the past are suppressed by the familiar exponential divergence of trajectories near the horizon. One such set of coordinates is obtained by applying an AdS$_2$ isometry (an $SL(2,\RR)$ transformation $\gamma$, defining $\tilde{t}=\gamma(t)$, $\tilde{x}^\pm=\gamma(x^\pm)$), a Rindler boost chosen such that $\tilde{t}=0$ corresponds to a proper time $u_0$ of order $k^{-1}$. In the new $\tilde{x}^\pm$ coordinates\footnote{Explicitly, we can choose $\frac{du}{d\tilde{t}}=1$ and $\frac{d^2u}{d\tilde{t}^2}=0$ at the time $t=t_0$, $u=u_0$, in which case we find \begin{equation*}
	\tilde{t}=\gamma(t)\sim \frac{1}{\pi \tilde{T}} \frac{t-t_0}{2t_\infty-t_0-t}
\end{equation*}}, the metric retains the same form \eqref{eq:xMetric}, and the dilaton profile is approximately given by the static black hole solution \eqref{eq:BHdilaton} with temperature $\tilde{T} = e^{-k u_0/2}T$, reflecting the fact that black hole is evolving adiabatically. This approximation to the dilaton profile is valid for a range of ingoing times of order $k^{-1}$, so in particular remains a good approximation a scrambling time ($\sim\log k^{-1}$) in the past, where the QES resides. In such coordinates, the QES lies near the would-be classical bifurcation surface at $\tilde{x}^+=\frac{1}{\pi\tilde{T}}$, $\tilde{x}^-=-\frac{1}{\pi\tilde{T}}$ of the comparison static spacetime, though the actual evolving spacetime lacks a classical extremal surface in this region since the approximation for the dilaton breaks down for sufficiently small $\tilde{x}^-+\frac{1}{\pi\tilde{T}}$. This perspective makes it particularly clear that the existence of the QES is rather universal, being insensitive to the history of the black hole after a few scrambling times.

The transition of the QES after the Page time is directly related to the growing gap between the left and right QESs, which we expect is a consequence of the growing entropy of the bath. Indeed, this expectation can be sharpened in the situation where our system can be well-approximated by a type I von Neumann algebra, the amount by which the right QES fails to be a left QES is related to the failure of any (state-dependent) map acting purely on the right CFT to approximate the state of the radiation. This failure can be directly attributed to the entanglement of the bath with the left CFT: the inaccuracy of the approximation is a consequence of the density matrix $\rho_{LB}$ not factorizing. One may speculate that the spacetime in between the two entanglement wedges may (in some appropriate sense) be emergent from entanglement with the bath.

Let us now discuss similar considerations in more general evaporating black holes. A model of this type closely related to our calculations above is given by starting with our thermofield double state and and some time turning on couplings to a pair of auxiliary baths $B_L$, $B_R$. Both $B_L,B_R$ begin in their ground state.  The coupling to $B_R$ is precisely as above and involves only our right system $R$, but we also introduce a corresponding coupling of $B_L$ to $R$.  Bulk causality then requires that the computation of the right quantum extremal surface is as before, and that the left quantum extremal surface behaves similarly.  The joint system $LR$ is a black hole that begins in a pure state (the thermofield double), radiates into $B_L$ and $B_R$, and initially increases in entropy.  In the semiclassical description of the bulk dual to $LR$ we see this increase in entropy through the increase of entropy of bulk quantum fields.  Indeed, the boundaries are homologous to the empty set, for which the generalized entropy at order $1/G_N$ is precisely the bulk entropy on a complete Cauchy slice through the bulk spacetime.

However, an interesting transition occurs if such radiation continues past the Page time, where the bulk entropy $S_{\mathrm{bulk}}$ on our Cauchy slice becomes greater than the density of states $S_{LR}$ of $LR$.  As this requires the original Bekenstein-Hawking entropy to exceed the ground state entropy by more than a factor of $2$, remaining in the near-extremal limit where our JT model is a controlled approximation to known dualities requires that this be achieved by repeatedly injecting a large number of low-entropy pulses of energy into the bulk and then waiting for the bulk to Hawking radiate this into the bath before sending in the next pulse, though one may alternatively study higher-dimensional AdS models where controlled dualities describe far-from-extremal black holes.  In either case, at late times we will find a new QES close to each horizon, since the relevant regions of the spacetime become adiabatically close stationary black holes so that we may again apply our earlier considerations.\footnote{One may also give an alternative argument using a maximin definition of the quantum extremal surface analogous to the classical HRT surface discussion in \cite{Wall:2012uf}; the arguments of \cite{Wall:2012uf} also apply to 1+1 quantum extremal surfaces so long as one assumes the quantum focusing conjecture of \cite{Bousso:2015mna} (though there are some subtleties in the choice of Cauchy slice used to define the entropy in the quantum focusing conjecture when the black hole is an open system).  Suppose that one turns off  the coupling at some late time, constructs a new spacetime from the appropriate late-time Cauchy data in the original spacetime, and studies $S_{\mathrm{gen}}$ for surfaces in the resulting new spacetime.  The new spacetime is essentially AdS-Schwarzschild outside the horizon, but has a long one-sided wormhole (with a large causal shadow) inside.  On any Cauchy surface, there will be {\it some} surface not too far inside the black hole and satisfying the homology constraint with area less than $4G_{N}S_{XL}$, and thus with generalized entropy also close to or smaller than $4G_{N}S_{XL}$.  In contrast, as above the empty set is associated with greater generalized entropy and so cannot be the maximim surface.  Instead, the maximin surface will must have generalized entropy bounded above by $4G_{N}S_{XL}$ up to logarithmic corrections.}.  Since energetic considerations will force $\frac{A}{4G_{N}} \approx S_{LR}$, beyond the Page time one finds $S_{\mathrm{bulk}} > \frac{A}{4G_{N}}$, so the quantum extremal surface with minimal generalized entropy is then near the horizon and is no longer the empty surface.  In this way the quantum extremal surface undergoes a first-order phase transition at the Page time.  Because the vast majority of the bulk entropy will be localized inside the black hole and local effects from quantum fields are small, the entropy of this second QES will be given by the horizon area $\frac{A}{4G_{N}}$ up to logarithmic corrections

Similar comments clearly apply to one-parameter families of so-called `bag-of-gold' spacetimes with large regions of spacetime behind a black hole horizon in which the bulk quantum fields are placed in a mixed state of large entropy.  For $S_{\mathrm{bulk}} < \frac{A}{4G_{N}}$, the minimal quantum extremal surface is the empty set, but for $S_{\mathrm{bulk}} > \frac{A}{4G_{N}}$ it jumps to near the black hole horizon; see \cite{Almheiri:2018xdw} for recent examples of this phenomenon in the context of the SYK model. In this context, both quantum extremal surfaces are in fact near classical extremal surfaces.    In both the small and large  $S_{\mathrm{bulk}}$ regimes, the dominant quantum extremal surface indicates that the entanglement wedge is associated with an entropy less than or equal to the density of states of the system to which it is dual.

Returning to the evaporating black hole, we see that tracking the quantum extremal surface using only perturbative semiclassical dynamics in the bulk fully reproduces the expected Page curve, including in particular the decay to zero of the entropy if the system decays to a non-degenerate ground state.  And it is also interesting that it does so by terminating the entanglement wedge within a short distance of the horizon of the remaining late-time black hole.

This observation will surely fan the flames of black hole information debates and on the possible role of firewalls \cite{Almheiri:2012rt,Almheiri:2013hfa,Marolf:2013dba} in particular.  On one hand, the termination of the entanglement wedge at the edge of the black hole may indicate that no meaningful spacetime exists farther in.  On the other, the fact that a purely perturbative semi-classical model of the bulk defines quantum extremal surfaces that reproduce the expected Page curve may indicate that the interior spacetime {\it is} meaningful, that no firewall is needed, and that the interior is somehow dual to the (arbitrary!) bath system (see e.g. \cite{Verlinde:2013qya,Maldacena:2013xja} for related ideas).  Indeed, if the bath were holographic and described by a bulk that connected to this interior by even a very small wormhole, this would be the natural conclusion of studying the QES for the bath.  While under normal circumstances the minimal QES would lie at the small wormhole (effectively the trivial surface discussed above), beyond the Page time the large bulk entropy in the bath system would move the QES out to the horizon.  In this context, for each boundary time the setting would be much like that of the original ER=EPR discussion \cite{Maldacena:2013xja} with similar potential implications for black hole information.

\begin{figure}
\begin{center}
\includegraphics[height=4cm]{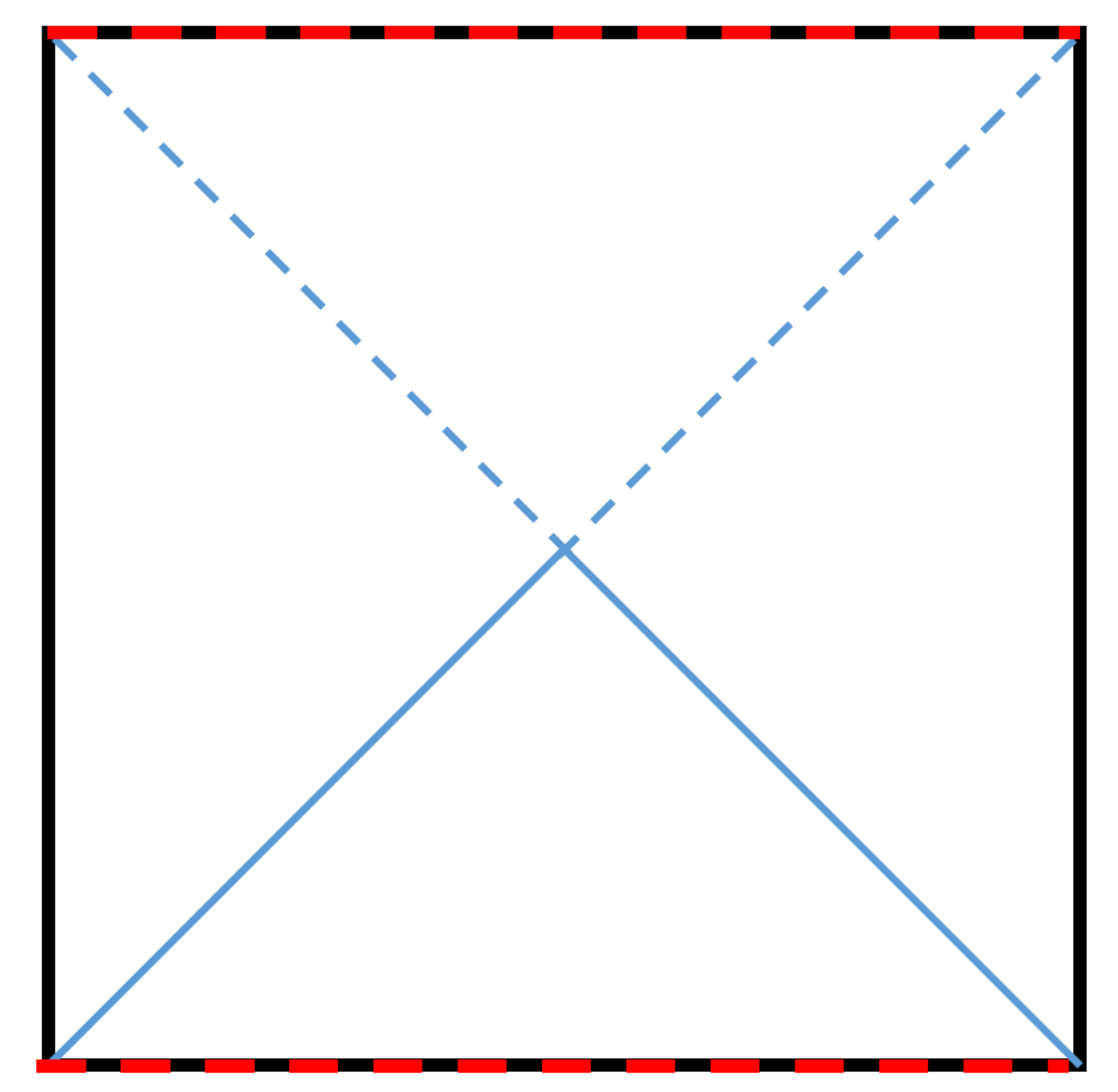}
\includegraphics[height=4.3cm]{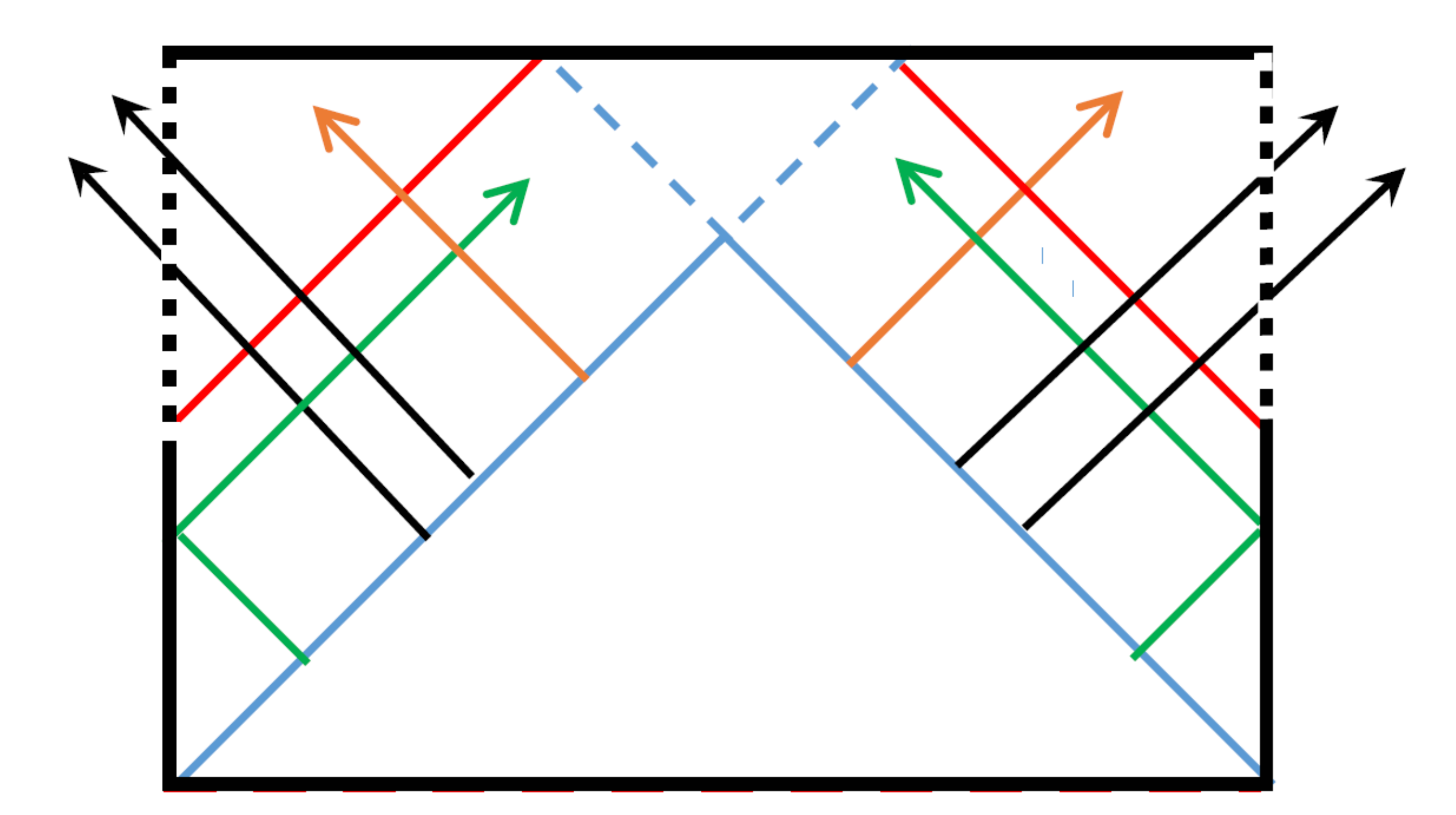}
\end{center}
\caption{(Left) For an eternal black hole dual to the thermofield double state, the segments of the horizons to the past of the bifurcation surface (solid lines) join to form a mostly-null Cauchy surface.  Data in this quantum state then propagates outward to the right and left.  (Right)  Turning on a coupling allows some modes from the above Cauchy surface to escape.  But those that reach the boundary before the coupling is turned on are reflected back into the singularity, and modes close to the original bifurcation surface are focused by an incoming pulse of positive energy (red lines without arrows) associated with turning the coupling on.  As a result of this focussing, semiclassical physics sates that such modes also fail to escape.}
\label{fig:PageTimeEvap2}
\end{figure}

However, such an interpretation would return us to the familiar problem that a perturbative semi-classical bulk description of the radiation does not provide the correlations between Hawking quanta necessary to purify the emitted radiation.  In particular, in the model described above one can use perturbative semi-classical bulk physics to track precisely the flow of entropy and information into $B_LB_R$ as was done for $B_R$ in section \ref{sec:QES}.  But since the modes on both right and left that escape into $B_L,B_R$ remain entangled with both modes that are reflected back into the black hole before the coupling is turned on and those that lie to the future of all modes that escape, doing so would find $B_LB_R$ to end in a highly mixed state; see figure \ref{fig:PageTimeEvap2}.  This is directly analogous to the Araki-Leib violation found for our one-sided evaporation at the end of section \ref{sec:QES} in the context where one executes repeated cycles of exciting the near-extremal JT-bulk by a small amount and then letting it Hawking radiate into the bath.

In both cases, unitarity thus requires that bulk semi-classical physics fails to correctly compute the late-time entropy of the bath.  A similar failure to correctly compute the
entropy of bulk quantum fields themselves is suggested by noting that in the full bulk spacetime the predicted entropy of such fields exceeds the density of states in the dual CFT.  Indeed, since this is precisely the feature that led to the QES phase transition at the Page time, it is interesting to ask if such a phase transition really occurs in a full non-perturbative treatment.  A plausible alternative speculation might be that the non-perturbative system instead evolves so as to become extremely close to this phase transition at late times -- perhaps even close enough that the concept of a definite entanglement wedge ceases to be well-defined.  On a positive note, however, since the entropy of semi-classical bulk fields within the entanglement wedge associated with the QES after the Page time appears consistent with the dual CFT density of states, it is at least self-consistent to use semi-classical physics within this wedge and to suppose that non-perturbative corrections become large only when one probes more deeply into the bulk.

We thus find that any perspective continues to lead to many open questions.
In order to make real progress in such debates it seems critical to understand more precisely what is meant by duality between field theory degrees of freedom and an entanglement wedge in the bulk.  Thinking of the entanglement wedge as defining the bulk region that can be reconstructed from the stated degrees of freedom suggests this be done by further investigating the role of quantum error correction and recovery maps in gravitational holography.  We thus look forward to further progress on this front, or on other related aspects of holographic duality.

\paragraph{ Acknowledgements} We wish to thank  D.~Harlow, A.~Harrow, Z.~Komargodski, N.~Lashkari, J.~Maldacena, and D.~Stanford.  AA is supported by funds from the Ministry of Presidential Affairs, UAE. N.E.~is supported by the Princeton University Gravity Initiative  and by NSF grant No. PHY-1620059.  D.M.~acknowledges support from NSF grant PHY1801805 and funds from the University of California.  Support for H.M.~was provided by a DeBenedictis Postdoctoral Fellowship and by funds from the University of California. H.M.~would like to thank C.~sinensis for moral support.

\bibliographystyle{jhep}
\bibliography{Union4}
\end{document}